\documentclass[lettersize,journal]{IEEEtran}
\usepackage{booktabs} 
\usepackage{caption}
\usepackage{adjustbox}
\usepackage{subcaption}
\usepackage{float}
\usepackage{url}
\usepackage{epstopdf}
\usepackage{graphics,graphicx}
\usepackage{cite}
\usepackage{balance}
\usepackage{array}
\usepackage{pifont}
\usepackage{multirow}
\usepackage{algpseudocode}
\usepackage{algorithm, algpseudocode}
\usepackage{amssymb}
\usepackage{amsmath,amsfonts}
\usepackage{algorithm, algpseudocode}
\usepackage{xcolor}
\usepackage{verbatim}
\usepackage{paralist, tabularx, enumitem}
\algrenewcommand\algorithmicrequire{\textbf{Input:}}
\algrenewcommand\algorithmicensure{\textbf{Output:}}

\begin{document}
	\title{FaiRIR: Mitigating Exposure Bias from Related Item Recommendations in Two-Sided Platforms}

	\author{\IEEEauthorblockN{Abhisek Dash, Abhijnan Chakraborty, Saptarshi Ghosh, Animesh Mukherjee, and Krishna P. Gummadi}
	\thanks{A. Dash, S. Ghosh and A, Mukherjee are with Indian Institute of Technology Kharagpur, Kharagpur 721302, India 
	(e-mail: dash.abhi93@iitkgp.ac.in; saptarshi@cse.iitkgp.ac.in; animeshm@cse.iitkgp.ac.in).}
	\thanks{A. Chakraborty is with Indian Institute of Technology Delhi, Delhi 110016, India. (e-mail: abhijnan@iitd.ac.in)
	}
	\thanks{K. P. Gummadi is with the Max Planck Institute for Software Systems, 66123 Saarbrucken, Germany. (e-mail: gummadi@mpi-sws.org )}
	}
	
	\maketitle	
	\begin{abstract}
		Related Item Recommendations (RIRs) are ubiquitous in most online platforms today, including e-commerce and content streaming sites. These recommendations not only help users compare items related to a given item, but also play a major role in bringing traffic to individual items, thus deciding the exposure that different items receive. 
		With a growing number of people depending on such platforms to earn their livelihood, it is important to understand whether different items are receiving their desired exposure.  
		To this end, our experiments on multiple real-world RIR datasets reveal that the existing RIR algorithms often result in very skewed exposure distribution of items, and the quality of items is not a plausible explanation for such skew in exposure.
		To mitigate this exposure bias, we introduce multiple flexible interventions (\textit{FaiRIR}) in the RIR pipeline. We instantiate these mechanisms with two well-known algorithms for constructing related item recommendations -- rating-SVD and item2vec -- and show on real-world data that our mechanisms allow for a fine-grained control on the exposure distribution, often at a small or no cost in terms of recommendation quality, measured in terms of relatedness and user satisfaction.\footnote{\textcolor{red}{This work has been accepted as a regular paper in IEEE Transactions on Computational Social Systems 2022 (IEEE TCSS'22).}}
	\end{abstract}
	\vspace{-2mm}
\section{Introduction} 
\label{sec:intro}
Recommendations are major drivers of traffic (and revenue) on two-sided market platforms, 
including e-commerce sites like Amazon or
Flipkart, and multimedia sites like YouTube, Spotify or 
Netflix~\cite{gomez2016netflix,smith2017two,AmazonRINSale}. 
There are two primary stakeholders in two-sided platforms: 
(1)~{\bf producers} (or sellers) of items (goods, contents or services) listed on the platforms, 
and (2)~their {\bf consumers} (or users). 
A few recent works have focused on the {\it fairness of recommendations} in such platforms, but mostly from the perspective of consumers~\cite{yao2017beyond,zhu2018fairness,edizel2019fairecsys,geyik2019fairness},  
such as, whether different groups of consumers experience similar 
quality of recommendations.

However, as the recommendations help consumers efficiently explore the item space, 
they also implicitly determine the amount of {\it exposure} different items get, 
affecting the revenues of their producers. For example, 
$30\%$ of Amazon's traffic originates from 
recommendations~\cite{sharma2015estimating}. Similarly, $80\%$ of movies 
watched on Netflix are driven by recommendations~\cite{gomez2016netflix}. 
With a large number of businesses and individuals depending on two-sided platforms 
to earn their livelihood~\cite{chakraborty2017fair}, there are growing concerns about 
the {\it fairness of these recommendations from the perspective of the producers or sellers}~\cite{patro2020fairrec,edizel2019fairecsys,yao2017beyond}.  
Even on the legislation front, a recent Indian regulation mandates e-commerce sites to treat their sellers fairly~\cite{pib2018ecommerce}. In this paper, we focus on the producer-side fairness considerations raised by recommendations.

\begin{figure}[tb]
	\begin{subfigure}{\columnwidth}
		\centering
		\includegraphics[width=\textwidth, height=3.25cm]{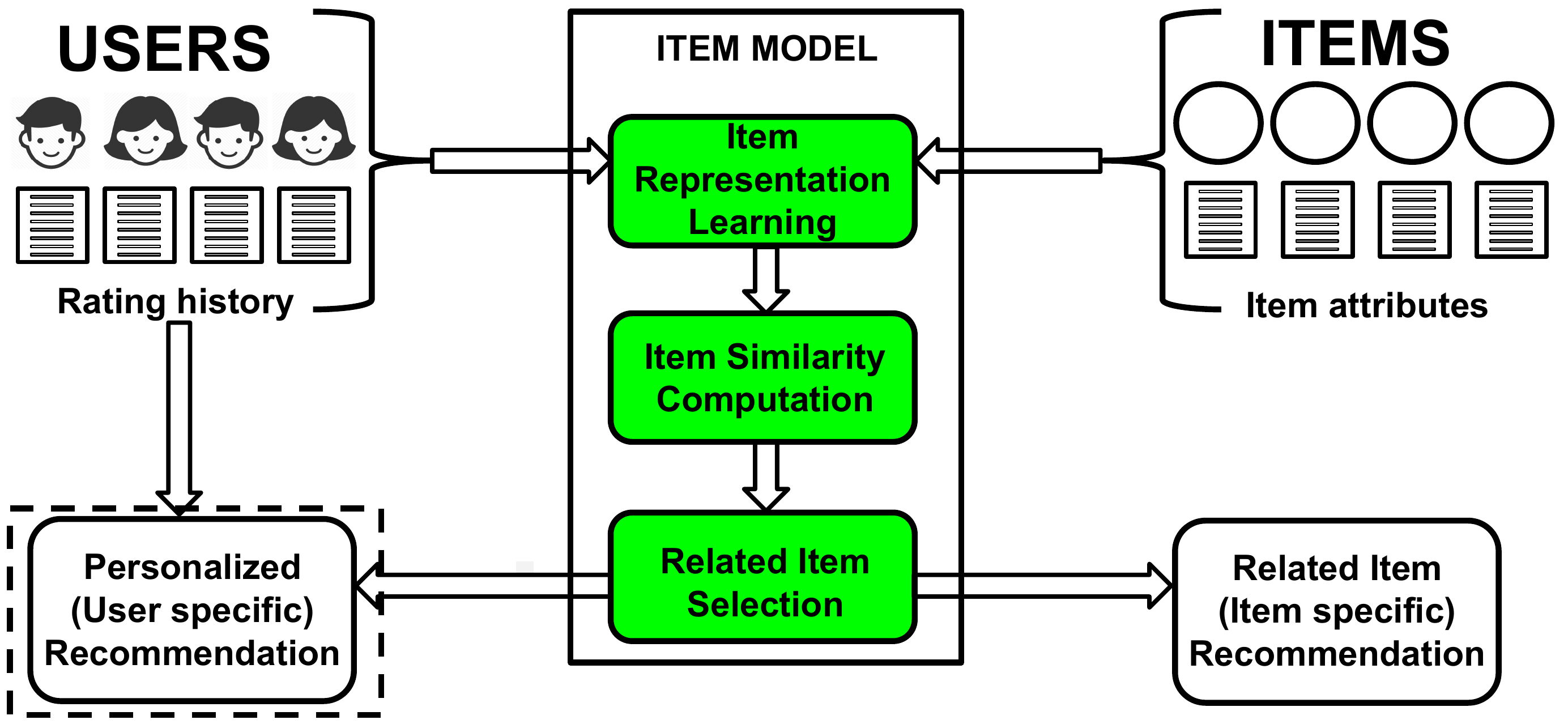}
		\label{fig:pipeline}
	\end{subfigure}%
	\vspace*{-5mm}
	\caption{{\bf A generic block diagram explaining item-based recommendation methods. Item models generated in these methods can be utilized for both related item recommendations and personalized recommendations. The latter, enclosed in broken rectangles in the diagram, is beyond the scope of the current work.}}
	\label{fig:recopipeline}
	\vspace*{-6mm}
\end{figure}

Recommendations in two-sided platforms are primarily of two types (see Figure~\ref{fig:recopipeline}): 
	(i)~item-specific {\bf Related Item Recommendations} (RIR),  
	e.g., `customers who viewed this item also viewed the following items'
	recommendations on Amazon, or `Up next' video recommendations on YouTube, and 
	(ii)~user-specific {\bf Personalized Recommendations},  
	e.g., `Related to items you've viewed', `Inspired by your shopping trends' recommendations on Amazon, `Because you watched X' on Netflix. 
	Notice while personalized recommendations are centered around the past interactions of a specific customer (to whom the recommendations will be shown), related item recommendations are centered in the context of a particular item~\cite{yao2018judging}. The underlying notion of relatedness can be of different types which will be discussed in detail in Section~\ref{sec:fairness}.
	Two recent works lately have looked at fairness for the producers~\cite{patro2020fairrec,patro2020incremental}, but they only consider the user-specific personalized recommendations. 
	To our knowledge, our work here is the first to investigate fairness issues in Related Item Recommendations (RIRs).

\vspace{1 mm} 
\noindent\textbf{Item exposure bias in RIRs}: 
As RIR algorithms recommend new items that are `related' or `similar' to the item currently being viewed by a user, there may arise situations where an item gets much more (or less) exposure than what it deserves.
For example, a poor-quality item may be recommended as related from a popular good-quality item (say, by virtue of being produced by the same manufacturer) and hence
the poor-quality item may end up getting much more exposure than it deserves. 
On the other hand, a good-quality item may fail to get
the desired exposure, simply because it is not recommended as directly related
to other popular items by a RIR algorithm. 
In fact, our investigation over real-world datasets (Section~\ref{sec: motivation}) shows that the relative exposure of items that would be induced by state-of-the-art 
RIR algorithms is often uncorrelated or disproportionate to the relative quality of the items. 
We term this discrepancy between the {\it observed} item exposure (as induced by RIRs) and the {\it desired} item exposure (e.g., based on item quality) as {\it exposure bias}. 

In this paper, we posit that by solely focusing on `relatedness' between items, RIRs may implicitly bias the exposure distribution of items in a manner that does not reflect a desired producer-fair exposure distribution.\footnote{Note that, exposure bias may get induced due to multiple explicit factors too, such as special relationship of certain items with the platform~\cite{dash2021umpire}; however such concerns are beyond the scope of the current work.}
We propose mechanisms to quantify and mitigate the exposure bias of RIRs. To formalise the concept of exposure bias, apart from the `observed' exposure of an item, we also need to have a notion of what is the `desired' exposure of that item. 
Since the notion of desired exposure is highly contextual, it cannot be riveted to a single operational definition. Therefore, we operationalize this notion of `desired' exposure in multiple ways allowing for multiple contextual assumptions. In fact, we develop the rest of our pipeline in such a way that {\it any new operationalization of the `desired' exposure can be seamlessly plugged in.}
We show that our proposed mechanisms can improve producer-side fairness of RIRs, with little or no impact on the utility of the recommendations to consumers.

\vspace{1mm}
\noindent
\textbf{Contributions}: We make the following contributions. 
\begin{itemize}[noitemsep,topsep=0pt,parsep=0pt,partopsep=0pt,leftmargin=*]
	\item We demonstrate the exposure bias induced due to two popular RIR algorithms -- rating-SVD and item2vec -- on two real-world datasets -- the MovieLens and Amazon product review datasets~\cite{harper2016movielens,he2016ups}. The choice of these two datasets is motivated by the two large businesses they represent -- the entertainment industry and the e-commerce industry.
	\item To counter the exposure bias, we propose {\bf FaiRIR}, a novel suit of three algorithms 
	applied at different stages in the RIR pipeline, that can minimize exposure bias
	while preserving the underlying relatedness of the recommended items to the best possible extent.
	\begin{compactitem}
		\item FaiRIR$_{rl}$, based on \textit{fair representation learning} 
		\item FaiRIR$_{sim}$, based on \textit{fair similarity computation} 
		\item FaiRIR$_{nbr}$, based on \textit{fair neighbor selection} 
	\end{compactitem}
	\item Extensive offline evaluations on the real-world datasets show that 
	FaiRIR can significantly reduce exposure bias, while preserving the underlying relatedness and utility of the recommendations toward the end-user.
	\item Finally, we conducted 
	a user survey on Amazon Mechanical Turk, which 
	further demonstrates the efficacy and utility of the proposed FaiRIR algorithms. 
\end{itemize}

\noindent We believe that the methodologies discussed in this paper are generic enough to be extended to different setups where related item recommendation algorithms are deployed. For better understandability and reproducibility, we have released our source codes at: \url{https://github.com/ad93/FaiRIR}.

	\vspace{-2mm}
\section{Background and Related Work}
We review prior works on related item recommendations, 
followed by 
the recent works on algorithmic fairness 
especially in the domain of recommendation systems.

\vspace{1mm}
\noindent \textbf{Related item recommendations:} 
RIR systems recommend items 
for a source item 
(typically, the item a user is consuming at a point in time) based on their relatedness to the source.
Evaluating relatedness between a pair of items is central to item-based collaborative filtering recommendations~\cite{desrosiers2011comprehensive,linden2003amazon}. 
Multiple prior works have proposed different approaches for identifying relatedness. 
For example, some approaches use user-rating matrices
~\cite{desrosiers2011comprehensive,sarwar2001item}, which may be factorized using matrix decomposition techniques like SVD~\cite{sarwar2000application}. 
Other approaches rely on implicit 
information such as clicks or co-purchases 
\cite{barkan2016item2vec,hu2008collaborative}. 

RIR algorithms are used in 
multiple two-sided platforms, with some platform-specific enhancements. 
For instance, Amazon uses RIRs in 
its product pages~\cite{linden2003amazon,smith2017two}, 
YouTube recommends videos through their `Up next' recommendations~\cite{covington2016deep}, 
Netflix uses RIR for recommending movies~\cite{gomez2016netflix}. 
In this work, we attempt to cover different types of relatedness 
considering each of the above categories, by examining 
user-item ratings (rating-SVD), and users' activity information (item2vec). 

\vspace{2mm}
\noindent \textbf{Algorithmic bias and fairness:} 
Recently, there have been extensive research on algorithmic bias and fairness across multiple disciplines 
~\cite{barocas2016big}. 
To 
tackle any inadvertent consequences of algorithmic decisions,
many recent studies have considered fairness from mainly two perspectives: 
(1)~{\it individual fairness}, which requires similarly deserving candidates should be treated similarly~\cite{zemel2013learning,lahoti2019ifair}, and 
(2)~{\it group fairness}, requiring different social salient groups should 
be treated similarly 
~\cite{pedreshi2008discrimination, pedreschi2009measuring}. 
While some studies focus on detecting
the discrimination (e.g.,~\cite{angwin2016machine,pedreshi2008discrimination}),
others suggest 
mitigation strategies by proposing fairness-aware
algorithms (e.g.,~\cite{lahoti2019ifair,zafar2017fairness,zehlike2017fa}).

Few recent works have considered group and individual fairness in {\it personalized recommendations}~\cite{yao2017beyond,edizel2019fairecsys,geyik2019fairness, patro2020incremental, patro2020fairrec} where the goal is to ensure 
that the recommendations do not discriminate against socially salient groups or individuals. 
To our knowledge, ours is the first attempt to consider {\it individual fairness in related item recommendations}, where we 
propose a novel algorithm {\it FaiRIR}, that attempts to provide the desired level of exposure to different items. 
	\vspace{-1mm}
\section{RIR Systems and exposure thereof} \label{sec:fairness}

In this section, we present the notion of relatedness between items and how we instantiate an item model capturing it. We also demonstrate the operationalization of exposure induced by an RIR algorithm with respect to the discussed instantiation. 

\vspace{-1mm}
\subsection{Relatedness of recommended items}
The primary goal of related item recommendations is to maximize the {\it relatedness of recommended items} to the source item
that the consumer has viewed/purchased/liked. 
Though there is no sacrosanct definition of relatedness, two items can be thought of as related over multiple dimensions: \\
(i)~{\bf Content based relatedness}, e.g., movies of the same genre, items from the same producer or brand, etc., \\
(ii)~{\bf Compatibility}: two items can be related if they are either the substitute or complement of one another, e.g., items that are frequently purchased together: a smartphone and its cover, \\ 
(iii)~{\bf External feedback on recommendation platforms}: user-actions such as likes and ratings, also define relatedness. For example, items being rated similarly, liked or disliked by a number of common users can be considered as related. 

\noindent
Relatedness, therefore, is subjective, and RIRs are judged 
based on whether the consumers find the source and the recommended items to be related. Additionally, the metric to measure relatedness between items is often domain-dependent.
{\it The concept of `relatedness' is analogous to `accuracy' or 'relevance' in the context of a related item recommendation system} -- just like classifiers are traditionally designed to optimize for accuracy, RIR systems are traditionally designed to optimize (maximize) relatedness.

\begin{figure}[tb] 
	\begin{tabular}{|c|c|}
		\hline
		Items & Related items \\
		\hline
		$I_1$  & $I_2$, $I_3$, $I_4$ \\
		\hline
		$I_2$  & $I_3$, $I_5$, $I_6$\\
		\hline
		$I_3$  & $I_4$, $I_5$, $I_6$ \\
		\hline
	\end{tabular}
	\adjustimage{width=0.75cm,valign=c}{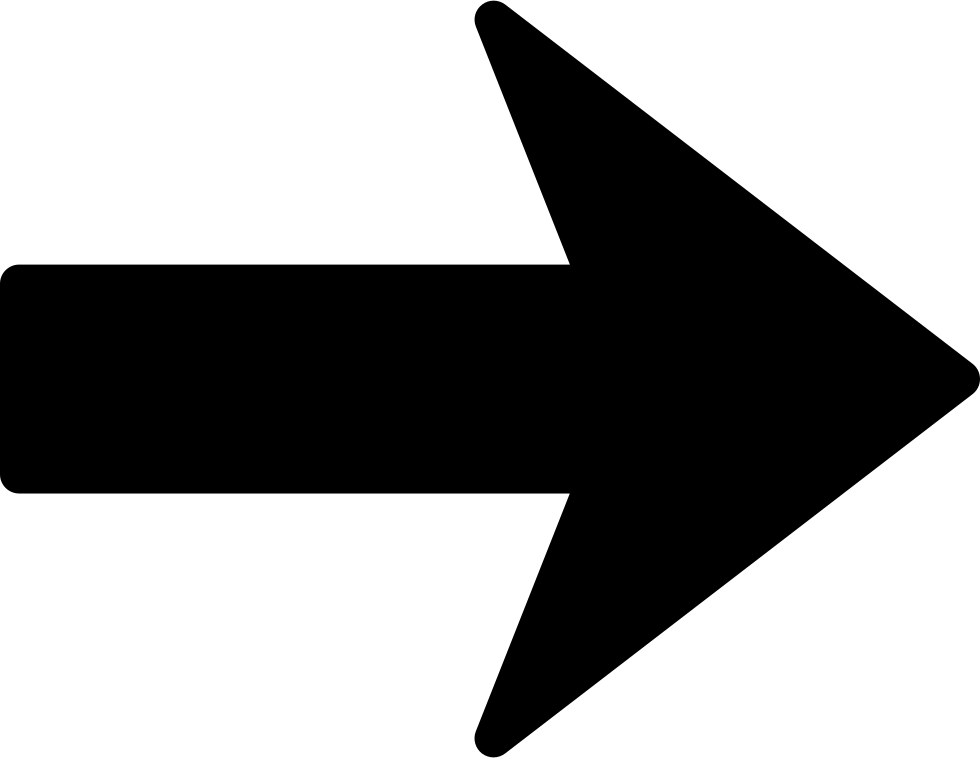}\quad
	\adjustimage{width=3.0cm,valign=c}{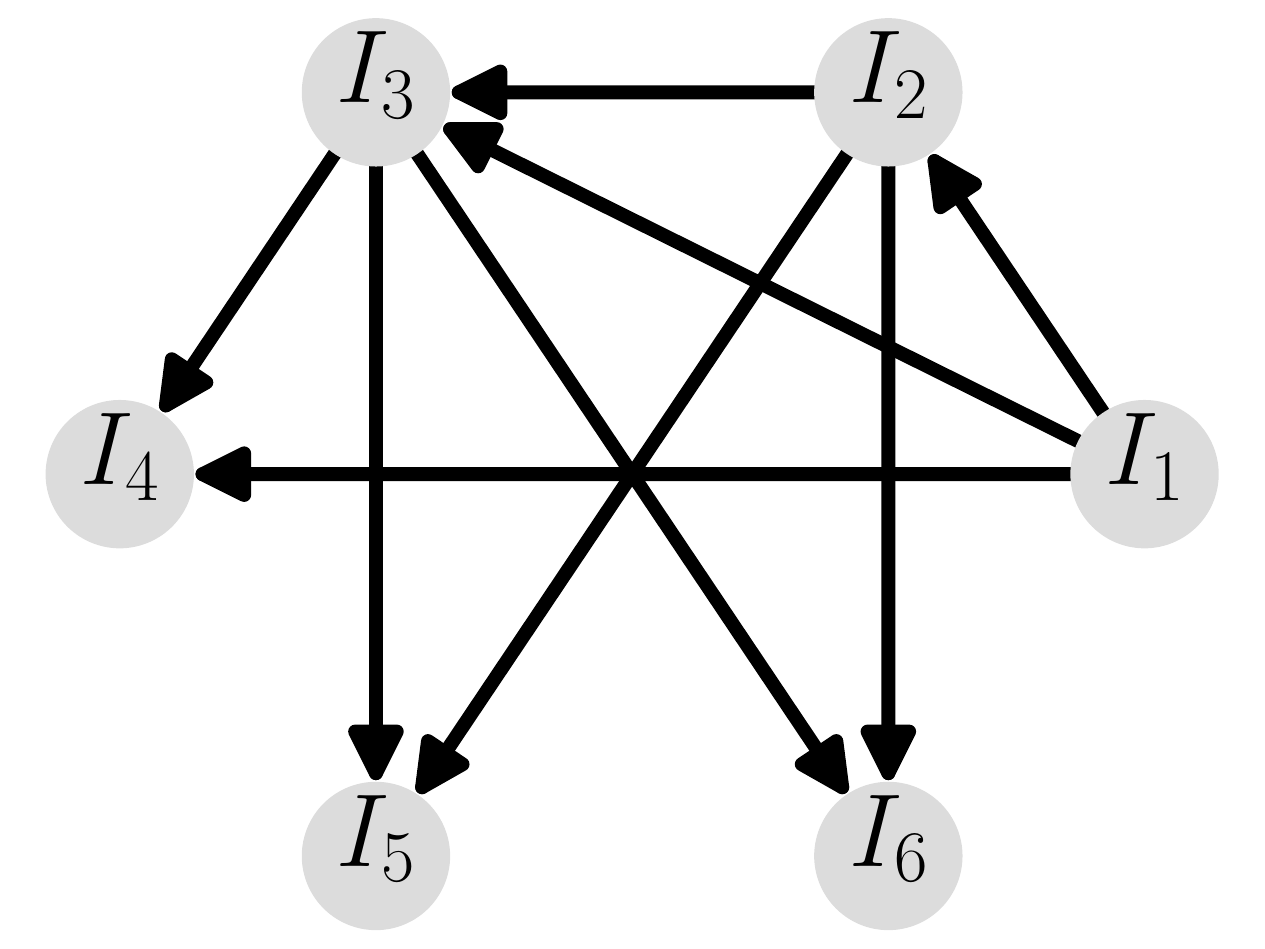}\quad
	\caption{\bf A sample item model and its corresponding RIN.}
	\label{Fig:RINCreation}
	\vspace{-7mm}
\end{figure}

\subsection{Instantiating Item Model by Related Item Network}

As shown in Figure~\ref{fig:recopipeline}, 
both Related Item Recommendations and Personalized Recommendation systems utilize an {\it Item Model} that captures the relatedness among items.
We now discuss an intuitive way to instantiate the Item Model, which was developed in our prior works~\cite{dash2021umpire, dash2019network}.

We utilize an instantiation of the Item Model of a recommendation system as a {\it Related Item Network} (RIN). 
A RIN is a directed network, with each node being analogous to an item in the universe,  
and a directed edge between two nodes implies that the corresponding source and destination items are related (based on some underlying notion of relatedness).
For instance, let us consider an item model as shown in the table in Figure~\ref{Fig:RINCreation}, and its corresponding RIN. 
Since item `$I_2$' is related to item `$I_1$', the corresponding nodes in the RIN are connected via a directed edge (from `$I_1$' to `$I_2$').

Once this instantiation of item model is constructed, a simple way to generate the Related Item Recommendations is as follows. For a particular source item, one can recommend those items to which it links in the RIN. For instance, in Figure~\ref{Fig:RINCreation}, the recommendations for source item `$I_1$' are items `$I_2$', `$I_3$', and `$I_4$'.

\subsection{Estimating observed exposure}\label{sec: EstOexp}

We define the observed exposure $E_o(i)$ of an item $i$ as the exposure it actually gets after the deployment of a RIR algorithm.
Ideally, the observed exposure of items should be quantified by click-through rates or other user interaction signals.
However, the availability of such comprehensive user-item interactions is seldom possible for third-party researchers due to the sensitive nature of the information.
Counting the number of recommendations received by items (analogous to in-degree of items in RIN) may be a possible work-around in such situations. 
However, the importance of all recommendations is {\it not} the same -- 
it varies with the source item, e.g., a recommendation from a popular source item is expected to yield more visibility (for the destination item) than a recommendation from a non-popular item.
 
Taking such observations into consideration, we use the `Random Surfer model'~\cite{random-surfer-model} to estimate the observed exposure. 
In general, users tend to visit the page of an item and then they start exploring different items recommended on the page. Alternatively, they can also randomly consume any other item thereafter. 
By simulating such user exploration for a large number of iterations, we take the {\it steady state visit frequency of a node} \textit{i} as its observed exposure $E_o(i)$ (more details can be found in our prior work~\cite{dash2021umpire}). Note that the notion of observed exposure of an item is very similar to PageRank of the corresponding node, in this formulation.

\vspace{1mm} \noindent
\textbf{Considerations during Random Surfer simulation: }
While simulating user browsing behavior, we note that different users can have different propensity to follow recommendations. 
The `teleportation probability' $\alpha \in [0, 1]$ of the Random Surfer model captures such considerations. 
The surfer chooses to traverse the recommended items with probability $(1 - \alpha)$, and teleport to a random item with probability $\alpha$. Throughout the paper, we report results for ($\alpha = 0.15$) which is the most prevalent value of teleportation in the literature~\cite{dash2021umpire, Brin98theanatomy}. Finally. we normalize the observed 
exposure scores of all items 
such that $\sum_{\forall i \in \mathbf{I}}^{}{E_o(i)} = 1$.
	
\section{Limitations of existing RIRs}
\label{sec: motivation}

Figure~\ref{fig:recopipeline} shows a schematic block diagram of the methods involved in building a model of {\it item-item relatedness}, (also called an \textit{item model}), which is then utilized
to make the recommendations at scale~\cite{smith2017two,nikolakopoulos2019recwalk}.
Item models 
capture the similarity 
between items based on either user-item interaction logs and/or item attributes~\cite{sarwar2001item, ning2011slim, desrosiers2011comprehensive}. From these logs, 
first a latent space representation of each item is obtained. 
Then, the similarity between pairs of items are computed from the latent representations, 
and $k$ most similar (related) items to a given item are generated. In case of RIRs, this item model is used to recommend items that are `related' or `similar' to an item.  
Next, we explore two popular RIR algorithms, 
and check whether their recommendations 
induce disproportionate exposure to items.

\vspace{-2 mm}
\subsection{Two popular RIR algorithms}\label{sec: RIRAlgos}
As a proof of concept, we consider the following two popular algorithms for generating related item recommendations -- (i)~rating-SVD, and (ii)~item2vec (detailed below). 
We choose these two algorithms because they cover some of the most common techniques for RIRs~\cite{yao2018judging}.
These algorithms take as input 
a {\it user-item rating matrix} $M$ whose $(i,j)$-th entry gives the rating that the user $i$ gave to the item $j$. 
First, these algorithms learn a latent space representation of different items (from $M$). 
To generate recommendations for a seed item $i$, the algorithms then generate top-$k$ neighbours of $i$, by ranking items based on their similarity with $i$ in the latent space.

\vspace{1 mm}
$\bullet$ \textbf{rating-SVD} applies Singular Value Decomposition (SVD)~\cite{sarwar2000application} to the user-item rating matrix $M$, and uses cosine-similarity for similarity evaluation
~\cite{sarwar2001item}. 
The underlying notion of relatedness between two items $i$ and $j$ can be described as `people who liked item $i$ are likely to like item $j$'. 
Following the setup in~\cite{yao2018judging}, we implement SVD with 128 dimensions. As a pre-processing step, we perform mean subtraction on the input and normalise each row of the final representations to a unit vector.

\vspace{1 mm}
$\bullet$ \textbf{item2vec}~\cite{barkan2016item2vec} is a replication of word2vec~\cite{mikolov2013distributed} representation learning
. item2vec substitutes items for words, and tries to find out co-occurrence patterns in user consumption. 
The underlying notion of relatedness among items can be defined as `people who consumed item $i$ are likely to consume item $j$ in close temporal proximity'. 
Following the setup in~\cite{yao2018judging}, we train the algorithm for 100 epochs by setting negative sampling to 15 and dimension of the output vector to 128. 

\subsection{Datasets for experiments} 
We performed our experiments on 
MovieLens datasets (1M and 10M)~\cite{harper2016movielens} and Amazon review dataset~\cite{he2016ups}, that are well-known benchmarks for recommendation tasks. 

\vspace{1 mm}
\noindent
\textbf{MovieLens datasets}: These datasets provide ratings and browsing logs of users;
the ratings come from real MovieLens users. 
The ratings range from $[0.5, 5]$ in half-star increments. 
Experiments on both the MovieLens 1M and 10M datasets yield qualitatively similar insights. 
Hence, we report results only on the MovieLens 10M dataset which contains $10,000,054$ ratings from $71,567 $ different users about $10,677$ distinct movies. 

\vspace{1 mm}
\noindent
\textbf{Amazon review dataset}: The Amazon product review dataset released by He \textit{et al.} \cite{he2016ups} comprises customer reviews and ratings for different Amazon products. For the purpose of this paper, we used the 5-core cellphone and accessories dataset. This dataset contains 194,439 reviews of 27,879 users for 10,429 different products, where each user and item has at least 5 reviews.

\vspace{1 mm}
\noindent
We applied the rating-SVD and item2vec algorithms on 
these datasets to find the top-$k$ RIRs for each item, and then created the RIN (see Figure~\ref{Fig:RINCreation}) 
to measure the observed exposures for each algorithm. 

\begin{figure}[tb]
	\centering
	\begin{subfigure}{0.43\columnwidth}
		\centering
		\includegraphics[width=\textwidth, height=3.0cm]{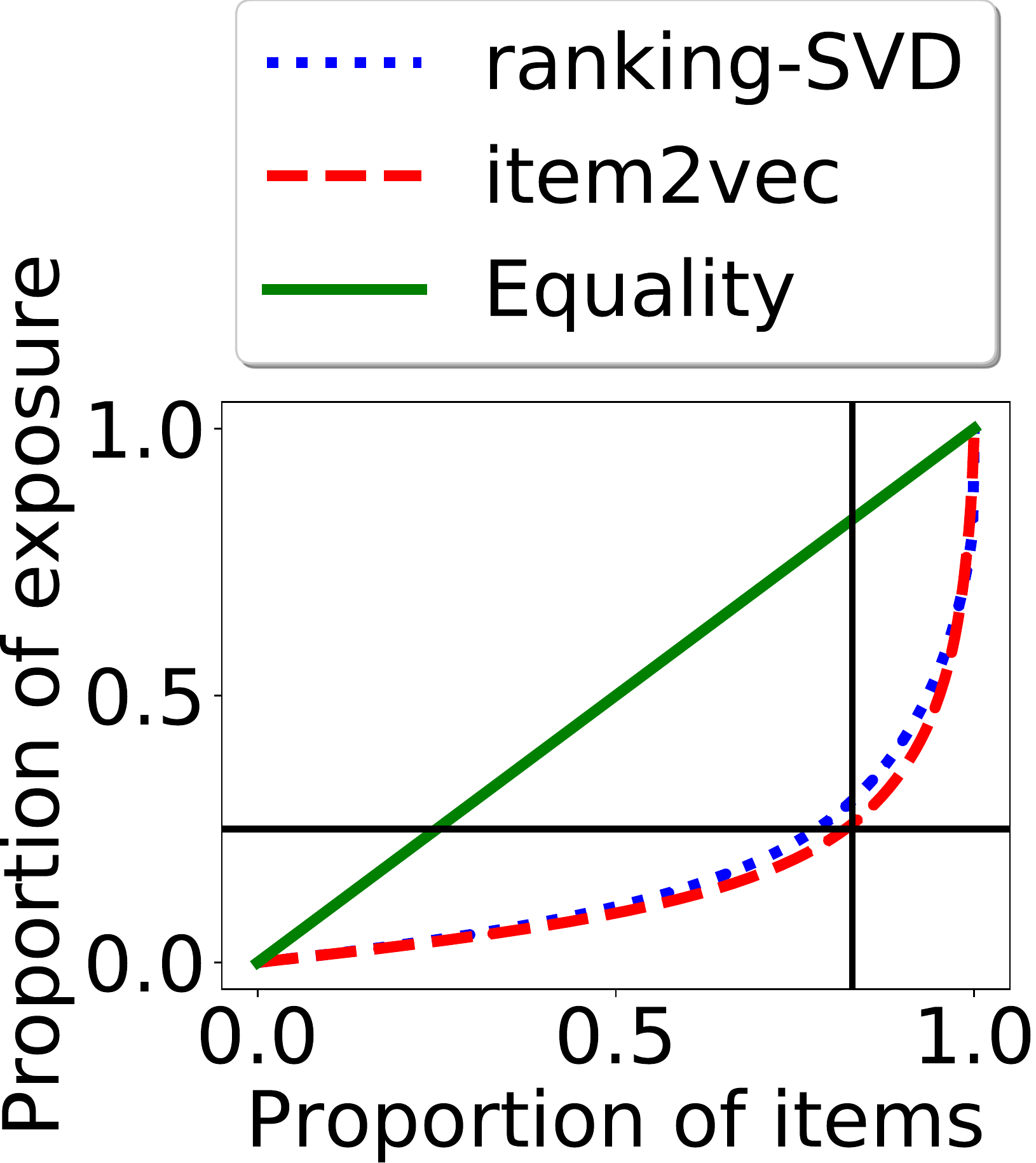}
		\caption{Amazon product review}
		\label{fig: AmazonExp}
	\end{subfigure}%
	\begin{subfigure}{0.43\columnwidth}
		\centering
		\includegraphics[width=\textwidth, height=3.0cm]{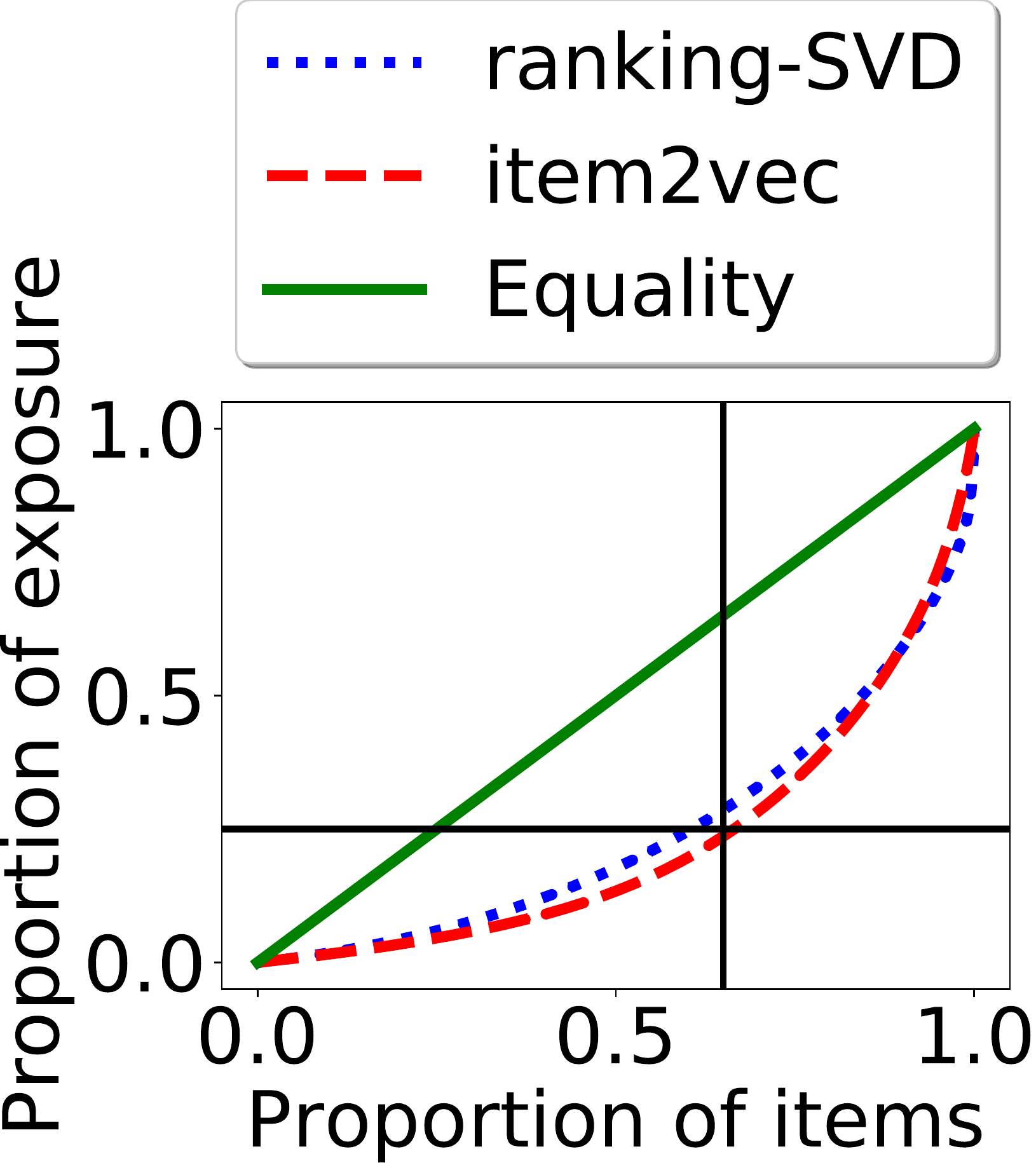}
		\caption{MovieLens }
		\label{fig: MLExp}
	\end{subfigure}
	
	\vspace*{-2mm}
	\caption{{\bf Lorenz plot showing the skew in observed exposure of items due to two state-of-the-art RIR algorithms over two real world recommendation datasets. 
	}}
	\label{fig: Lorenz}
	\vspace*{-5mm}
\end{figure}

\subsection{Skew in exposure of items due to RIRs}
\label{sub:existing-rir-bias}
We applied rating-SVD and item2vec algorithms on MovieLens and Amazon product review datasets (described above) to find the top-$k$ recommendations for each item. We experimented with different $k$ values, and the results were qualitatively similar in all cases. Hence, we report the results for the top-$10$ related item recommendations.

The existing RIR algorithms perform very well as regards to finding related items for items. However, the exposure that different items get is found to be heavily skewed. Figure~\ref{fig: Lorenz} shows the cumulative proportion of exposure distribution on the $y$-axis, and $x-$axis shows the cumulative proportion of items in the increasing order of their exposure (from left to right). The horizontal line (in black) corresponds to 25\% of the entire exposure and the corresponding vertical line denotes the percentage of items accounting for it.
Figure~\ref{fig: Lorenz}(a) shows that the top $25\%$ of the items with most exposure in the Amazon cellphone and accessories dataset account for $75\%$ of the entire exposure. Thus, a small fraction of the item-set gets very high fraction of the entire exposure; put differently, there is very low item-space coverage.

\vspace{1 mm}
\noindent
\textbf{Are the top items deserving of the exposure?} Based on the observation of this skewed exposure distribution, a plausible question can be raised: are those $25 \%$ items of very high quality? If so, one may argue that such items, having better quality than others, deserve more exposure than others, i.e., gap in quality can explain this skew in the exposure distribution. To further dwell on this particular line of argument, we investigated both the quality distribution and exposure distribution together.

There can be many measures for quality,
based on domain experts' opinion (e.g., critical reviews of movies), public opinion (e.g., ratings given by consumers on e-commerce sites), awards won by movies, and so on. Specifically, in this work, we assume the {\it average user-rating} of an item as the quantification of its quality. We further normalize the quality distribution so that the quality of all items add up to $1$. Note that we understand the limitations of using average user ratings as a quality measure; however, we believe that such ratings in most cases are a reflection of an item's perceived quality for a given user.

\noindent
\textbf{Observations}: A striking observation we made is the large gap between the quality and exposure of different items. 
Only 6\% -- 7\% of the total number of items in the Amazon dataset have comparable quality and exposure, with a stark disparity between quality and exposure for a large majority of items. 
\begin{figure}[tb]
	\centering
	\begin{subfigure}{0.45\columnwidth}
		\centering
		\includegraphics[width=\textwidth, height=3.0cm]{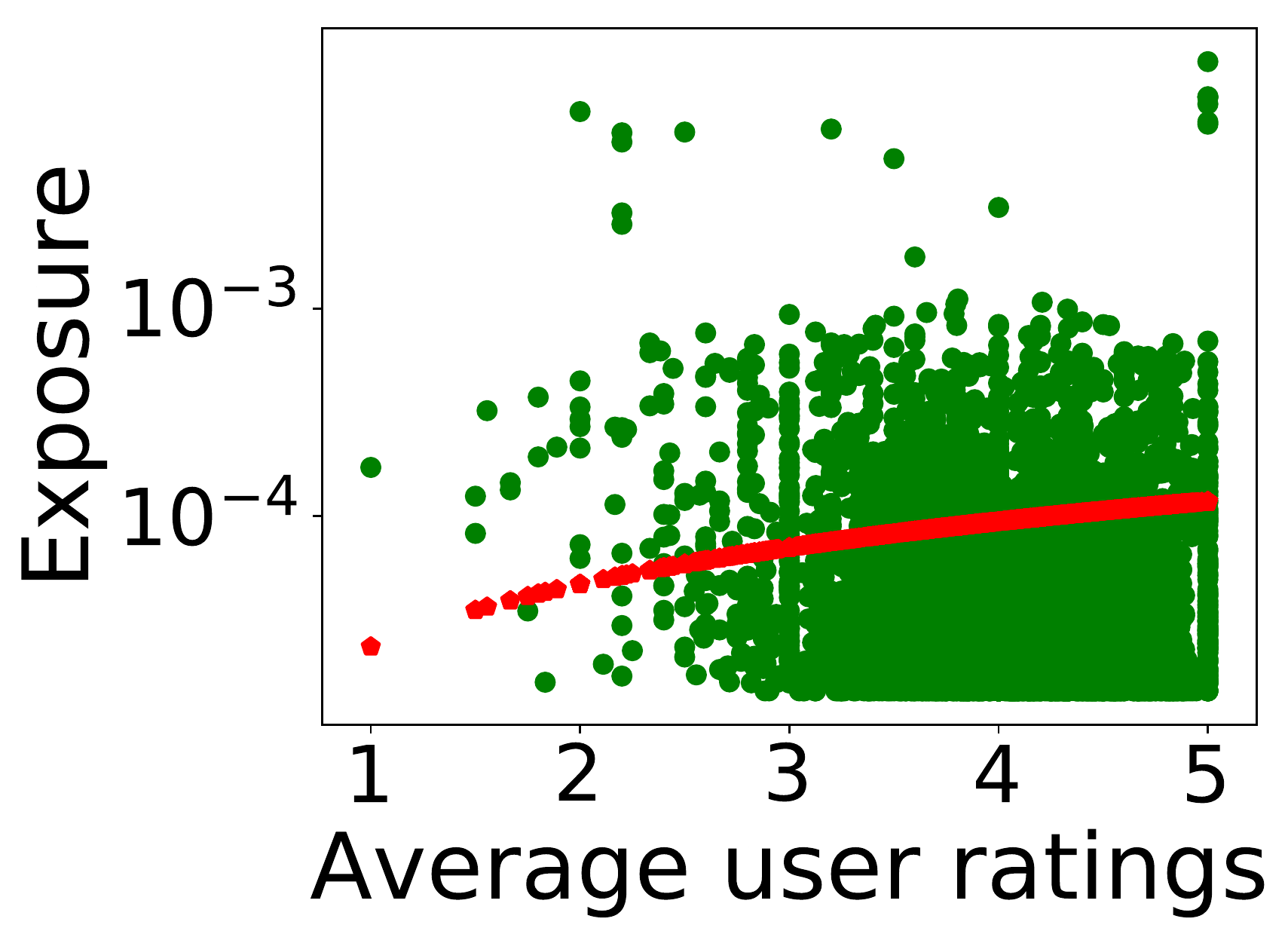}
		\caption{Amazon product review}
		\label{fig: IMDbExp}
	\end{subfigure}%
	\begin{subfigure}{0.45\columnwidth}
		\centering
		\includegraphics[width=\textwidth, height=3.0cm]{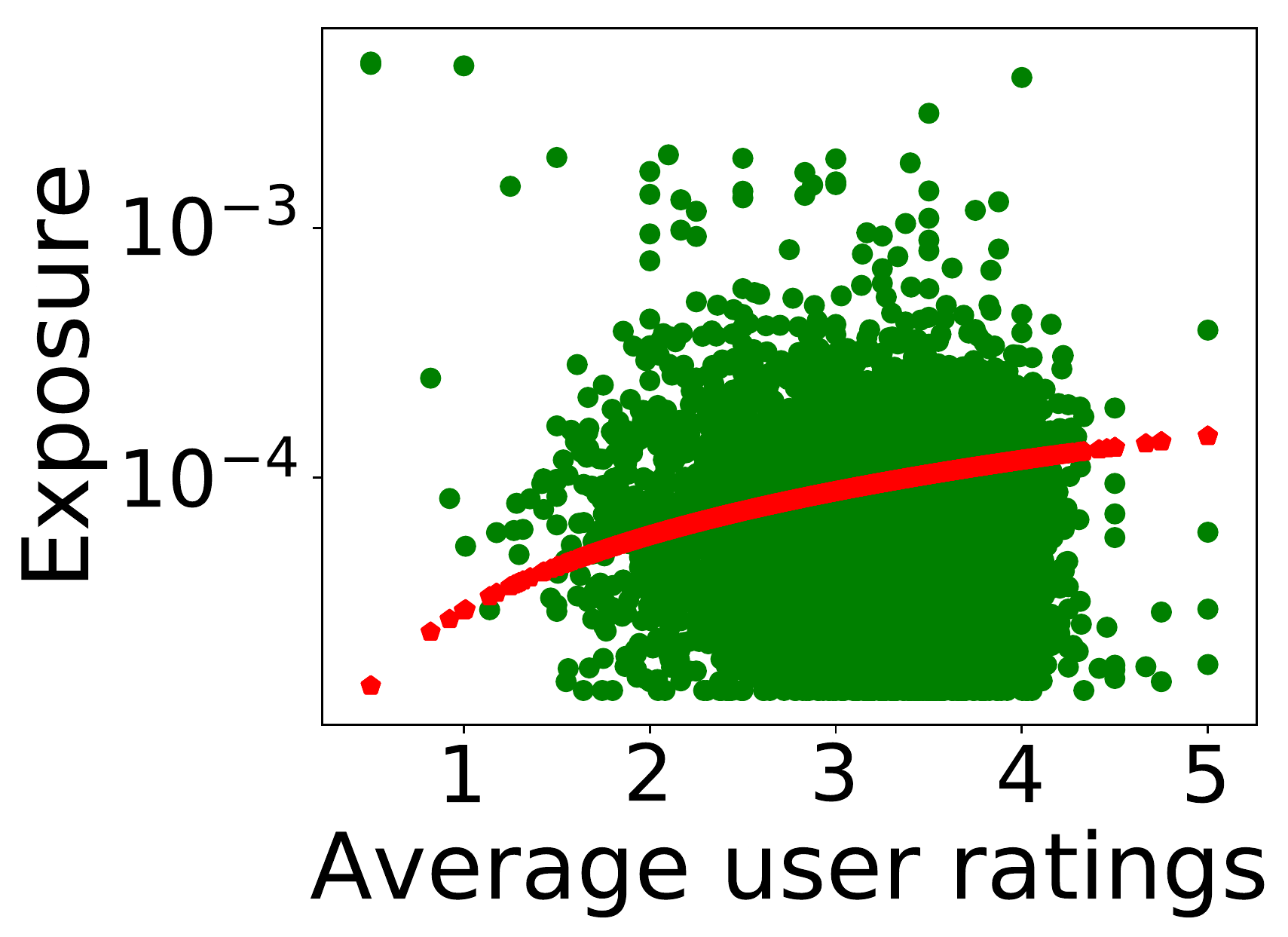}
		\caption{MovieLens }
		\label{fig: SVDExp}
	\end{subfigure}
	
	\vspace*{-2mm}
	\caption{{\bf (color online) Scatter plot of log of observed exposure and average user rating of different items in Amazon product review, and MovieLens dataset (using rating-SVD). The red curve shows the log plot of the quality distribution.}}
	\label{fig: Exp}
	\vspace*{-5 mm}
\end{figure}
Figure~\ref{fig: Exp} shows scatter plots for ranking-SVD RIRs on Amazon review (Figure~\ref{fig: Exp}(a)) and MovieLens (Figure~\ref{fig: Exp}(b)) datasets. 
In these figures, the $x$-axis is the average user rating (quality) of different items and $y$-axis is the log of their observed exposures. The red curve in each figure is the log plot of the quality distribution, while the green dots show the observed exposure of the corresponding items. Note that many items having low user-ratings (in the range $[1, 2]$) enjoy significant observed exposure, while many high quality items are deprived of the same. Similar trends are found for item2vec on both the datasets (results omitted for brevity).

\begin{table}[tb]
	\noindent
	\scriptsize
	\centering
	\begin{tabular}{ |p{1.3cm}|p{2.5cm}|p{4.0cm}| }
		\hline
		{\bf RIN} & {\bf High-quality items} & {\bf Low-quality recommended items} \\
		\hline
		\multirow{2}{*}{Amazon } & Plantronics Windsmart Headset (4.3) & Plantronics Headset Charging Cradle for Voyager (2.2) \\
		\cline{2-3}
		& Motorola H385 Bluetooth Headset (5.0) & Motorola Hands-Free Headset (2.9) \\
		\hline
		\multirow{2}{*}{MovieLens} & Rear Window (8.5) & Meatballs part II (3.7)\\
		\cline{2-3}
		& 12 Angry Men (8.9) & Stop! Or My Mom Will Shoot (4.2) \\
		\hline
	\end{tabular}	
	\caption{{\bf Examples of low-quality items that get over-exposed being recommended by high-quality items. Values within () are average user-ratings.}}
	\label{Tab: examplesRIN}
	\vspace{-3 mm}
\end{table}

We observe that RIR links often lead to poor-quality movies from high-quality ones,
resulting in higher exposure of unworthy items, potentially at the cost of other more worthy items. The reason behind such phenomenon can be formation of recommendation links (due to the underlying relatedness algorithm). Note that if a poor-quality item gets recommended on the web-page of a popular high-quality item, it may be bestowed some exposure simply due to the merit of the high quality item from which it is getting recommended. For instance, many customers will organically end up at the item page of the high quality item and in turn, they well get recommended to the said low quality items increasing their accessibility and visibility. This phenomenon may get reinforced over time as many customers would end up exploring both the items in the same session of exploration~\cite{dash2021umpire, dash2019network}.

Table~\ref{Tab: examplesRIN} shows a few illustrative examples of the above phenomenon.
For example, a high-quality item `Motorola H385 Bluetooth Headset' (avg. user rating 4.9) recommends the relatively low-quality item `Motorola Hands-free Headset' (avg. user rating 2.9) in both rating-SVD and item2vec recommendations (note that there is a quality drop of almost $40\%$). 
Similarly, for the MovieLens dataset, several high-quality movies having user-ratings higher than $8.0$ recommend movies of much lower user ratings (according to ranking-SVD). Note that, the average user ratings for movies are obtained from their corresponding IMDb pages.

Table~\ref{Tab: examplesDistortion} shows a few illustrative examples of the amount of distortion in exposure of some items introduced by RIR algorithms. 
Many high quality items, e.g., `Samsung Galaxy S5 SM-G900H', etc. have been severely deprived of exposure at the cost of significant exposure of items like `Google Nexus Wireless Charger', etc.

Given the huge inventory size of online platforms, RIR systems are one of the primary tools through which users explore the universe of items. Thus, these systems play a significant role in deciding 
how much 
{\it exposure} (or visibility) different items receive. Hence, an unjust exposure distribution can have detrimental repercussions for the producers of these items and their livelihood. Being keyed to relatedness, these algorithms inadvertently overlook these important aspects of different stakeholders in the business cycle. In section~\ref{sec: mitigation} we suggest three different intervention mechanisms (\textbf{FaiRIR}) to appropriately adjust the skew in the exposure of items.

\begin{table}[tb]
	\noindent
	\scriptsize
	\centering
	\begin{tabular}{ |p {4 cm}|p {4 cm}| }
		\hline
		\bf High ratings, deprived of exposure & \bf Low ratings, with high exposure \\
		\hline
		\multicolumn{2}{|c|}{\textbf{Amazon Cellphones and Accessories}}\\
		\hline
		Samsung 3.5mm Stereo Headset for Galaxy S4 (4.2, $\frac{1}{10}$ times)& Google Nexus Wireless Charger from Google (1.2, 10 times)\\
		\hline
		TechPro Galaxy S4 ShatterProof Premium Tempered Glass Screen Protector (4.83, $\frac{1}{8}$ times) & Sony Ericsson Liveview watch micro display for Android Devices (1.4, 9 times)\\
		\hline
		Samsung Galaxy S5 SM-G900H (4.8, $\frac{1}{7} $ times) & HTC One Screen Protector, Spigen Steinheil GLAS.t SLIM Premium Tempered Glass (1.35, 9 times) \\
		\hline
		\multicolumn{2}{|c|}{\textbf{MovieLens}}\\
		\hline
		Bittersweet Motel (8.0, $\frac{1}{11} $ times) & Diebinnen (5.5, $17$ times)\\
		\hline
		Men With Guns (7.6, $\frac{1}{10} $ times) & Slaughterhouse Rock (3.9, $10$ times)\\
		\hline
		Braveheart (8.3, $\frac{1}{12} $ times) & 
		Get Over it (5.7, $9$ times)\\
		\hline		
	\end{tabular}	
	\caption{{\bf High-quality (low-quality) items that are differently exposed. Values within () denote average user-ratings and the ratio of observed exposure to normalized quality.}}
	\label{Tab: examplesDistortion}
	\vspace{-8 mm}
\end{table}
	\section{Desired Exposure and Exposure bias}
Next, we discuss how exposure can be 
{\it fairly distributed} among a set of items, by motivating it through the lens of distributive justice~\cite{yaari1984dividing}. We 
then define `Exposure bias', given the desired and observed distributions of exposure. 

\vspace{-2 mm}
\subsection{Desired exposure of items}
\label{sec:desired}
Exposure in an online platform is a beneficial commodity, 
hence the producers of items would prefer having more of it (than having less). In such a scenario, an intuitive notion of  fairness would be \textbf{equality of exposure}, i.e., the exposure should be uniformly distributed among all the items (by recommended them uniformly). 
However, the characteristics of the 
items 
should also be taken into account, since these characteristics may provide prima facie grounds for a departure from equality~\cite{yaari1984dividing}. 
For instance, all items are probably {\it not} of similar merit or intrinsic quality. This difference in `merit' or `quality' can be a justified reason for departure from equality. Thereby, the `desired exposure' of an individual (item) can be determined by its `deservingness' (merit)\footnote{Desiredness should not be confused with deservingness, i.e., desiredness $\ne$ deservingness in general. Deservingness, in contrast, is an extreme case of desiredness.}. This departure from equality is well established through the notion of \textbf{meritocratic fairness} and the related literature on meritocracy~\cite{joseph2018meritocratic,joseph2016fairness}. For instance, a high-quality item is considered more deserving of 
user attention than a low-quality item. 

Alternatively, the desired exposures of various items can also be driven by a broader idea of societal welfare. 
For instance, YouTube `Up next' related video recommendation has recently been criticised for leading users to far-right echo chamber and extremist content~\cite{youtubeRadical}, potentially influencing elections (e.g., the Brazilian presidential election~\cite{youtubeBrazil}). In response, YouTube tweaked its `Up next' algorithm, and started recommending Fox News videos from far-right conspiracy theory videos, instead of other videos from the same channels~\cite{youtubeFoxnews}. Clearly, YouTube deemed some videos {\it unworthy} of the exposure they were getting earlier and decided to nudge users to follow other videos. 
In some scenarios, 
it might be legally required to provide each item with some minimum amount of exposure, regardless of the item attributes~\cite{patro2020incremental,patro2020fairrec}.

\vspace{1 mm}
\noindent
\textbf{Desiredness as a control knob for fairness: }Note that, we do {\it not} argue for any particular notion of desired exposure distribution; rather, the formulation and algorithms given in the subsequent sections are \textit{\textbf{agnostic}} to any measure of desiredness. 
Rather than advocating for any specific desired exposure, we perceive desiredness as a necessary {\it controllable knob} in our framework to ensure fairness in the final outcomes. 
Hence, if some legislation or a particular platform has a sacrosanct quantification of the desiredness of each item, the same can be easily plugged into our proposed fairness interventions.

\vspace{-2mm}
\subsection{Estimating desired exposure}
We denote the desired exposure of item $i$ as $E_d(i)$, and the desired exposure distribution over all items as $E_d$.
In this work, as a proof of concept, we consider a generic formulation to accommodate multiple types of desired exposure distributions. We consider, a fraction $\beta \in [0, 1]$ of the total exposure is equally distributed among all items. This fraction of the exposure takes care of the minimum exposure of all items (and their producers). It is meant to provide all items with some minimum exposure to satisfy the basic needs of the items and their producers (as argued in~\cite{patro2020incremental}). The remaining ($1-\beta$) fraction of the total exposure is distributed proportional to the quality or merit of individual items, thus advocating \textit{meritocratic fairness}~\cite{joseph2018meritocratic,joseph2016fairness}. Notice, the above formulation of $E_d$ reduces to purely meritocratic distribution of exposure for $\beta = 0$, and to uniform distribution of exposure for $\beta = 1$. The exposure distributions are normalized so that the total exposure of all items in the item-set sum up to 1, i.e., $\sum_{i \in \mathbf{I}}^{}{E_d(i)} = 1$.

As mentioned earlier, in this work, we assume the {\it average user-rating} of an item as the quantification of its merit / quality. The importance that we attach to an item's merit to obtain its desired exposure is controlled by the parameter $\beta$.

\vspace{1mm}
\noindent
{\bf A potential limitation of user-ratings:} One potential concern about using average user-ratings as a quality metric, might be that the number of ratings an item gets is partly driven by the 
existing recommendation algorithms. 
However, we believe that, although a user may have been led to an item via some recommendation, her rating would reflect the inherent quality of the item as perceived by her. 
Further, we also considered a slightly different quality measure -- average user-rating of an item, weighted by its number of ratings. The qualitative results of the analyses remained similar
in this setting too. 
Hence, for simplicity and completeness we 
consider the average user-rating score to be the indicator of quality throughout this paper. 

\vspace{-1mm}
\subsection{Defining exposure bias}  \label{sub:exp-bias}
According to our formulation, a RIR system would be fair (unbiased), if it gives every item an observed exposure that is proportional to its desired exposure. 
Since $E_o(i)$ and $E_d(i)$ denotes the {\it observed and desired exposures} of item $i$, mathematically, a RIR system is fair if $\frac{E_o(i)}{E_d(i)} = \frac{E_o(j)}{E_d(j)} \hspace{2mm}  \forall {i, j} \in \mathbf{I}$.
As discussed in the previous section, a RIR system $R$ may lead to items getting different observed exposures than what is desired. 
{\it Exposure Bias} ($ExpBias$) is the 
deviation caused due to $R$ between the desired and observed exposure of items.
Following the set up in our prior work~\cite{dash2021umpire}, we measure $ExpBias$ by KL divergence~\cite{cha2007comprehensive} between the observed exposure distribution $E_o = \{E_o(i) \; \forall i \in \mathbf{I}\}$ and the desired exposure distribution $E_d = \{E_d(i) \; \forall i \in \mathbf{I}\}$: 
\setlength{\belowdisplayskip}{0pt} 
\setlength{\abovedisplayskip}{0pt} 
\small
\begin{align}
ExpBias(R) = D_{KL} (E_o||E_d)= \sum_{i }{E_o(i) \hspace{1mm} log \hspace{1mm} \Big(\frac{E_o(i)}{E_d(i)}\Big)}
\end{align}\normalsize

\vspace{1mm}
\noindent
\textbf{Categorization of items}:
Based on the observed and desired exposure of items, we categorize items in three different classes based on how closely the observed exposure replicates their desired exposure.\\
(a) {\bf Under-exposed:} item $i$ is \textit{under-exposed} if $1-\epsilon \leq \frac{E_o(i)}{E_d(i)}$, \\
(b) {\bf Over-exposed:} item $i$ is \textit{over-exposed} if $\frac{E_o(i)}{E_d(i)} \geq 1+\epsilon$, \\
(c) {\bf Adequately-exposed:} item $i$ is \textit{Adequately-exposed} if $1-\epsilon \leq \frac{E_o(i)}{E_d(i)} \leq 1+\epsilon$, \\
While this threshold ($\epsilon$) can be chosen based on prier context and established regulations, in this paper, we use $\epsilon=0.2$. Note that similar thresholds have been used in multiple prior works too~\cite{chakraborty2017who, dash2018beyond, dash2021umpire}. 

	\vspace{-2 mm}
\section{Mitigating Exposure Bias}
\label{sec: mitigation}

\begin{figure}[tb]
	\begin{subfigure}{\columnwidth}
		\centering
		\includegraphics[width=\textwidth, height=3.25cm]{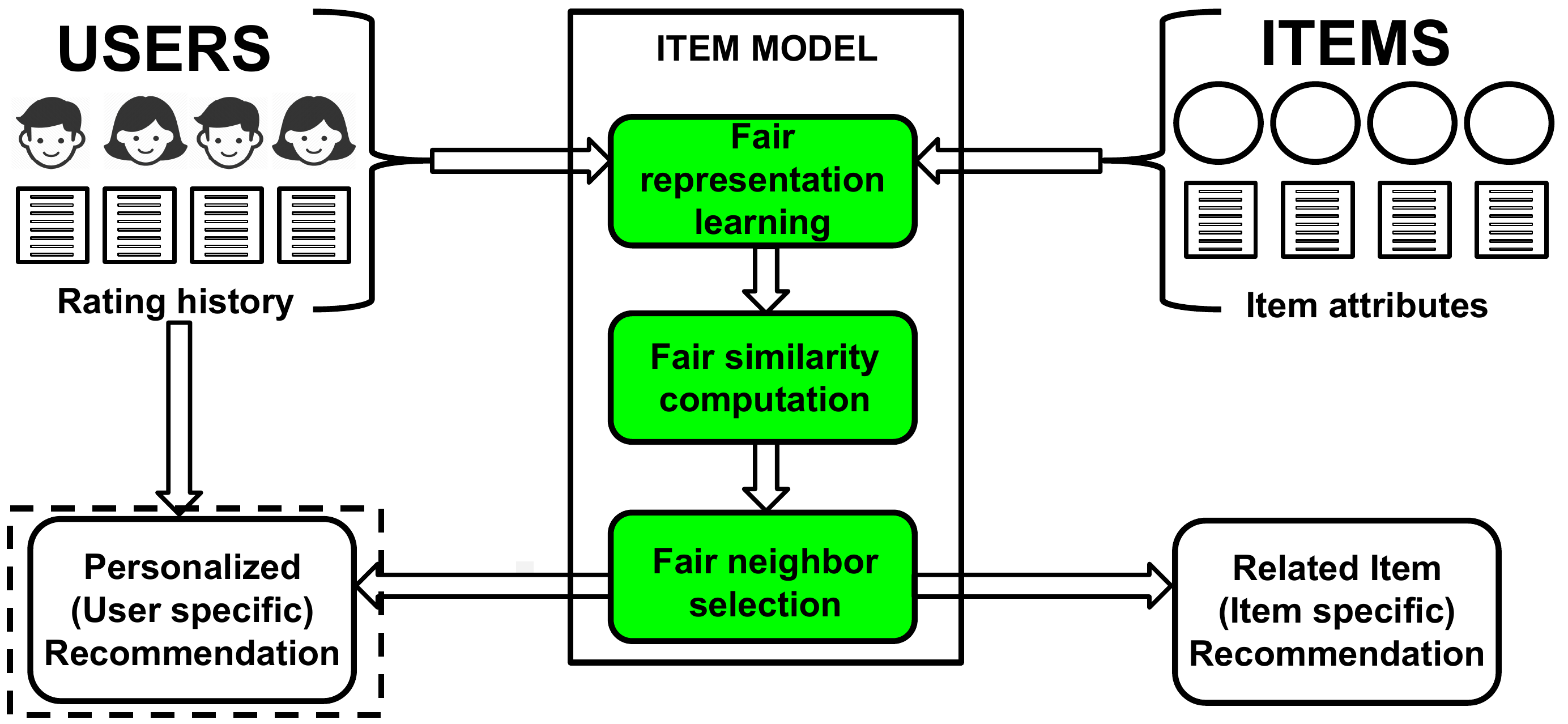}
		\label{fig:pipeline}
	\end{subfigure}%
	\vspace*{-5mm}
	\caption{{\bf Recommendation pipeline shown earlier in Figure~\ref{fig:recopipeline}, now shown with FaiRIR suit of fairness interventions.}}
	\label{fig:Fairecopipeline}
	\vspace*{-5mm}
\end{figure}

In this section, 
we propose multiple interventions (\textbf{FaiRIR}) in the recommendation pipeline (shown in Figure~\ref{fig:Fairecopipeline}), that can 
reduce exposure bias, 
by making exposure of item $i$ proportional to its desired exposure 
$E_d(i)$, while maintaining the relatedness of recommendations. \\  
(I)~{\bf FaiRIR$_{rl}$ (Fair representation learning)}: Change the latent space representation of items such that 
items with similar desired exposure come closer in the latent space. \\ 
(II)~{\bf FaiRIR$_{sim}$ (Fair similarity computation)}: Introduce desired exposure based similarity into the metric to compute similarity of items.\\
(III)~{\bf FaiRIR$_{nbr}$ (Fair neighbor selection)}: For a particular source item, instead of selecting the most similar items, 
select items considering both similarity and desired exposure. \\
Next, we will elaborate these three procedures.

\subsection{\textbf{FaiRIR$_{rl}$}: Fair representation learning} \label{sec: Phase 1}

Our first approach tries to learn fair representations of the items from the 
latent space representations obtained from 
vanilla rating-SVD or item2vec, to mitigate the induced exposure bias.
This approach is inspired by 
recent works attempting to incorporate individual fairness in various algorithms~\cite{zemel2013learning, lahoti2019ifair, lahoti2019operationalizing}.
Similar to~\cite{lahoti2019ifair}, we learn 
representation that preserves the similarity of desired exposure between items while minimizing the loss in relatedness, and, thereby, reconciling both desired exposure and relatedness. 

\vspace{1 mm}
\noindent
\textbf{Input}: Let us assume there are $M$ items, and for each item, we have $N$ latent attributes from the representations learned by the existing 
RIR algorithms (e.g., rating-SVD, item2vec), resulting in 
a $M \times N$ matrix $X$. We denote the record for the $i$-th item 
as vector $x_i$ and the value $x_{ir}$ denotes the $r$-th attribute of $x_i$. 
Note that, this representation space only captures 
the relatedness between 
items.

\vspace{1 mm}
\noindent
\textbf{Considering desired exposure}:  
We design a \textit{desiredness graph} $D_G(V, E)$, where every node represents an item, i.e., $V=\mathbf{I}$, and every item is connected to $k$ other items who have similar desired exposure, i.e., $E = \{(i,j) : E_d(i)\simeq E_d(j)\}$. 
We obtain the node representation ${x}_{i}^{*}$ by using a network representation learning algorithm node2vec~\cite{grover2016node2vec} for each item $i$. Note that, representation ${x}_{i}^{*}$ of an item captures similarity of the items based on their desired exposure only.

\vspace{1 mm}
\noindent
\textbf{Goal}: Our goal is to learn fair representation $\tilde{x}_i$ for item $i$, which 
preserves both relatedness and similarity in desired exposure.
Analogous to the input, the output is again a matrix $\tilde{X}$ with dimension $M \times N$ where $i$-th row represents the fair representation of item $i$.

\vspace{1 mm}
\noindent
\textbf{A simple approach to achieve fair representations:}
Given the two sets of representations $x_i$ and $x^*_i$, the most intuitive way to generate a fair representation is to concatenate both the sets of representations. We denote such an approach as FaiRIR$_{concat}$. 
Such a simple approach may be effective because while $x_i$ encodes the information regarding relatedness, $x^*_i$ encodes the information regarding desiredness. 
However, as we shall show later in Section~\ref{sub:expt-mitigate-bias}, the performance of such a simple approach is not stable across different datasets and vanilla RIR algorithms. Hence, we proceed to learn the fair representations by optimizing a loss function which reconciles between relatedness loss and desiredness loss in the final learnt representations (described next).

\vspace{1 mm}
\noindent
\textbf{Probabilistic clustering}: Following prior works~\cite{zemel2013learning, lahoti2019ifair}, our framework treats the goal of computing fair representation as the formal problem of probabilistic clustering. The aim is to learn $\mathbb{K}$ prototype vectors $v_h$ $ (\forall h \in \{1, 2, ..., \mathbb{K}\})$, such that item $i$ $(\forall i \in \{1, 2, ..., M\})$ is 
assigned to clusters in a probabilistic manner, such that the probabilities encode the distance between item $i$ and the prototype $v_h$. 
Given the distance function $d$ in an $N$-dimensional latent space, the probability that item $i$ belongs to cluster with prototype $v_h$ is given as: 
$
\label{eqn: softmax}
u_{i h} = \frac{exp(-d(x_i, v_h))}{\sum_{t=1}^{\mathbb{K}}exp(-d(x_i, v_t))}
$.
Notice such probabilistic clustering based set up can be viewed as a low-rank representation of the input matrix $X$ with $\mathbb{K} < M$, so that we are able to reduce the attribute values into a more compact form. 

\noindent
\textbf{Output representation}: The fair representation $\tilde{X}$, a matrix of dimension  $M \times N$ of fair output vectors $\tilde{x}_i$ ordered row-wise, includes
(a)~$\mathbb{K}<M$ prototype vectors $v_h$ of $N$ dimensions.
(b)~A probability distribution $u_i$, of $\mathbb{K}$ dimensions, for each item $i$. $u_{ih}$ represents the probability that item $i$ belongs to the cluster with 
prototype $v_h$. Mathematically,
$
\tilde{x}_i = \sum_{h \in \{1,...\mathbb{K}\}}{u_{ih}*x_{j}}
$, 
where $u_{ih}$ is as defined earlier. 

\vspace{1 mm}
\noindent
\textbf{Loss function}: Next, we present the loss function which optimizes for the reconstruction loss between the input representation $X$ and the fair output representation $\tilde{X}$ while preserving the desiredness of the products.
\scriptsize
\begin{align}
\label{eqn: optimize}
L = \underbrace{\lambda * \sum_{i=1}^{M}\sum_{r=1}^{N} (x_{ir}-\tilde{x}_{ir})^2}_{\textrm{Relatedness loss}} + \overbrace{\mu * \sum_{i, j \in \{1,..., M\}}[d(\tilde{x}_i, \tilde{x}_j)-d(x_i^*, x_j^*)]^2}^{\textrm{Desired exposure based similarity loss}}
\end{align}
\normalsize
This loss $L$ has two separate parts:
(i)~\textbf{Relatedness loss:} the sum of the squared errors between the input representation matrix $X$ and the (low dimensional) output representation matrix $\tilde{X}$, and
(ii)~\textbf{Desired exposure based similarity loss:} 
$d()$ captures the distance between desired exposure of two items, computed as \textbf{Euclidean distance} between their vector representations.
Hyper-parameters $\lambda$ and $\mu$ decide the importance we want to associate with these two losses. 

\subsection{\textbf{FaiRIR$_{sim}$}: Fair similarity computation} \label{sec: Phase 2}

As discussed in section~\ref{sec: motivation}, the existing cosine similarity between two item representations, 
$
sim(x_i, x_j) = \frac{x_i.x_j}{\lvert\lvert x_i \rvert\rvert\lvert\lvert x_j \rvert\rvert}
$,
accounts for relatedness; however, it does {\it not} account for the relative gap between their desired exposure. 
In 
this intervention, we propose to incorporate similarity of desired exposure between items along with relatedness. 
If $E_d(i)$ and $E_d(j)$ denote the desired exposure of items $i$ and $j$ respectively, then the new similarity measure is defined as:
\begin{align}
\label{eqn: p2}
sim(x_i, x_j) = \underbrace{exp(-\lvert E_d(i)-E_d(j)\rvert)}_{\textrm{Desired exposure based similarity}}*\overbrace{\frac{x_i.x_j}{\lvert\lvert x_i \rvert\rvert\lvert\lvert x_j \rvert\rvert}}^{\textrm{Relatedness similarity}}
\end{align}
\normalsize
\noindent 
Using the similarity metric mentioned in equation~\ref{eqn: p2}, we promote items having higher desired exposure based similarity and relatedness
by giving them higher similarity score. 

\subsection{\textbf{FaiRIR$_{nbr}$}: Fair neighbour selection}  \label{sec: Phase 3}

Next, instead of changing the representation or the similarity metric, 
we change the way 
the items are selected for recommendation. 
In practice, against every item, an equal number (say, $k$) of items are recommended; these $k$ items are usually most similar to the source item, based on some similarity metric. 
However, 
the number of recommendations each item will end up receiving is not controlled. 
We propose a fair way of 
selecting the $k$ neighbours 
such that the likelihood of an item being selected is proportional to its desired exposure.
That is, the recommendation would be fair if the likelihood of a highly desired item being recommended 
is greater than that of a less desirable item. If $R_t$ denotes the list of items recommended for item $t$, then for all item-pairs $(i, j)$,
$\forall t \in \{1,2,..., M\}, P(i\in R_t) \geq P(j\in R_t) | E_d(i) \geq E_d(j)
$
\normalsize
If an RIR algorithm follows the above equation
, while preserving the notion of relatedness, it is likely to 
mitigate exposure bias.

\begin{algorithm}[t]
	\scriptsize
	\begin{algorithmic}
		\Require $Desired$: number of recommendations desired by each item, $k$: number of recommendations per item and similarities among all items $Sim$
		\Ensure $Recommendation$
		
		\Function{Find\_Neighbor}{$u, k, Desired, Sim$}
		\State {$Similarity$ = Ranked list of items based on their similarity to item $u$} 
		\State {$Desiredness$ = Ranked list of items based on their Desired number of recommendations} 
		\State {$Final$ = Aggregated ranked list using $Similarity$ and $Desiredness$}
		\State {Return top-$k$ items based on $Final$ ranked list}
		\EndFunction
		\\
		\Procedure {Recommendation}{$k, Desired, Sim$}
		\State {$Recommendation = \phi$}
		\ForAll{item $u$ in the item space}
		\State {RelatedItems = \Call {Find\_Neighbor}{$u, k, Desired, Sim$}}
		\ForAll{item $v$ in RelatedItems}
		\State {$Recommendation = Recommendation \cup {(u, v)}$}
		\State {$Desired[v] = Desired[v]-1$}
		\If {$Desired[v] = 0$}
		\State {remove $v$ from $Desired$}
		\EndIf
		\EndFor
		\EndFor
		\State {Return $Recommendation$}
		\EndProcedure
		\caption{Fair neighbor selection}
		\label{Algo}
		\vspace{-1mm}
	\end{algorithmic}
\end{algorithm}

Algorithm~\ref{Algo} details our proposed algorithm. 
Through the $Desired$ dictionary, it ensures that the likelihood of recommendation among different items follow the similar distribution as defined by their desired exposure. Effectively, for any source item, we have two ranked list of items according to their relatedness and desired exposure. In order to reconcile between these two rankings, 
we use a well known rank aggregation method, based on borda count~\cite{borda1784memoire}. Intuitively, any item having higher rank in both the ranked lists will be considered the most suitable related item to be recommended. 

\vspace{1 mm}
\noindent
\textbf{Practical applicability of FaiRIR}:
There are two key questions regarding the practical applicability of FaiRIR algorithms in real-world setting -- 
(i)~does the deployment need a knowledge of the internal details of the RIR algorithm being used?, and
(ii)~can the FaiRIR algorithms adapt to a dynamic setting, e.g., when new items emerge intermittently.

FaiRIR$_{rl}$ and FaiRIR$_{sim}$ can be used only when internal details of the RIR algorithm are known 
(e.g., user-item logs, similarity metric etc.). Also, given that these interventions are applied in the earlier stage of the recommendation pipeline, they are more suitable to adapt to a dynamic setting.
FaiRIR$_{nbr}$, on the other hand, can be used even by considering the related item recommendation
algorithm as a black box, provided we have the recommendation outputs and the desired exposure of all items. 
For example, in case of an existing online platform, we do not exactly know the 
RIR algorithm that is used. However, since we can capture the recommendation outputs in the form of an RIN, we can use any network level similarity as a proxy for relatedness and rewire the RIN to reduce the exposure bias.

To adapt to the dynamic setting, in case of FaiRIR$_{nbr}$, one can start recommending the newly emerging items from items which already exist in the network and are similar to them (where the similarity can be 
based on some metadata or content-based measures, such as the genre or actors of movies); thereby mitigating the requirement of re-wiring the entire RIN. 

	\vspace{-2mm}
\section{Experimental Evaluation} \label{sec:evaluation}
We evaluate our interventions (FaiRIR algorithms) on 
MovieLens and Amazon Cell Phone and Accesories datasets. 
The algorithms are evaluated based on: (1)~their effectiveness in mitigating exposure bias (Sec~\ref{sub:expt-mitigate-bias}), (2)~the relatedness of their recommendations (Sec~\ref{sub:expt-reco-relatedness}), and (3)~the overall utility  of the recommendations to the end-users (Sec~\ref{sec: user_survey}). 

\begin{table}[tb]
	\noindent
	\scriptsize
	\centering
	\begin{tabular}{|p{0.85cm}|p{1.2cm}||p{1.0cm}|p{1.0 cm}|p{1.0cm}||p{1.0 cm}| }
		\hline
		Algorithm &  & Over & Adequate & Under & $ExpBias$ \\
		\hline
		& Vanilla 	& 25.89 \% & 10.95 \% & 63.16 \% & 0.71\\ \cline{2-6}
		& FaiRIR$_{concat}$	& 19.77 \% & 9.71 \% & 70.52 \% & 0.67\\ \cline{2-6}
		rating-& FaiRIR$_{rl}$	& 34.69 \% & 27.4 \% & 37.91 \% & 0.15\\ \cline{2-6}
		 -SVD& FaiRIR$_{sim}$	& 24.3 \% & 15.00 \% & 60.7 \% & 0.39\\ \cline{2-6}
		& FaiRIR$_{nbr}$	& 0.07 \% & 98.9 \% & 0.03 \% & 0.003\\ \cline{2-6}
		\hline \hline
		& Vanilla 		& 26.75 \%& 12.38 \% & 60.87 \% & 0.56  \\ \cline{2-6}
		& FaiRIR$_{concat}$	& 22.95 \%& 19.65 \% & 57.4 \% & 0.35  \\ \cline{2-6}
		item2vec & FaiRIR$_{rl}$	& 31.06 \%& 34.19 \% & 34.75 \% & 0.10  \\ \cline{2-6}
		& FaiRIR$_{sim}$	& 31.74 \%& 23.43 \% & 44.83 \% & 0.21  \\ \cline{2-6}
		& FaiRIR$_{nbr}$	& 0.04 \% & 99.94 \% & 0.02 \% & 0.002  \\ \cline{2-6}
		\hline 
	\end{tabular}
	\caption{{\bf \% of movies that are over, adequately and under-exposed in MovieLens dataset (vanilla and 
			intervened rating-SVD \& item2vec) with desired distribution proportional to quality of the movies (i.e., $\beta = 0.0$).}}
	\label{Tab: ML-10M}
	\vspace{-2mm}
\end{table}

\vspace{-2 mm}
\subsection{Mitigation of exposure bias} \label{sub:expt-mitigate-bias}


\noindent
\textbf{FaiRIR$_{rl}$}: We applied FaiRIR$_{rl}$ on the learnt representations of items (from the vanilla rating-SVD and item2vec algorithms) over the MovieLens and Amazon datasets, with the following parameter settings.

\noindent
\underline{Parameter setting}: We initialize the prototype vectors $v_j$ to random values from uniform distribution in $(0, 1)$. To account for the variations due to the initialization, we report the results obtained from the best of three runs. 
For the hyperparameters $\lambda, \mu$ in Eqn.~\ref{eqn: optimize}, we performed a grid search over the set \{0.01, 0.1, 1.0, 10, 100\}. For $K$, we performed a grid search over the set \{10, 20, 30\}. 
We found the best performance (argmin for the loss function in eqn~\ref{eqn: optimize}) for $\lambda = 1, \mu = 0.01$ and $K = 20$.

\noindent
\underline{Results}: The effectiveness of FaiRIR$_{rl}$ is shown in Table~\ref{Tab: ML-10M} (for Movielens) and Table~\ref{Tab: Amazon} (for Amazon) over both rating-SVD and item2vec algorithms.  
Compared to the original algorithms, the exposure bias has decreased significantly, with an increase in fraction of items being adequately exposed. E.g., for rating-SVD, percentage of items adequately exposed has increased from $06.62\%$ to $23.71\%$, and the exposure bias has reduced from $1.28$ to $0.18$ for the Amazon dataset. 

\noindent
\textbf{FaiRIR$_{rl}$ outperforms FaiRIR$_{concat}$}: Recall in Section~\ref{sec: mitigation}, we discussed two potential ways for fair representation learning. While one was the aforementioned optimization framework, the other was simple concatenation of representations learnt from vanilla RIR ($x_i$) algorithm and desiredness graph ($x^*_i$). In Tables~\ref{Tab: ML-10M} and~\ref{Tab: Amazon}, we show the efficacy of the approach in mitigating exposure bias for $\beta = 0.0$. While using FaiRIR$_{concat}$ we see reasonable improvement on both the datasets for representations learnt from Item2Vec approach; the performance was not so great for representations learnt from SVD approach. In either case, the proposed optimization based representation learning (FaiRIR$_{rl}$) outperforms the concatenation approach and its performance is more robust across datasets and across RIR algorithms. Hence, in the remainder of the paper, we shall consider FaiRIR$_{rl}$ to be the fair representation learning based mitigation approach and compare it with interventions at other stages of the pipeline.

\noindent
\textbf{FaiRIR$_{sim}$}: The effect of FaiRIR$_{sim}$ is also shown in Tables~\ref{Tab: ML-10M} and~\ref{Tab: Amazon}. 
In all cases, the $ExpBias$ has decreased, with an increase in percentage of items being adequately exposed.

\noindent
\textbf{FaiRIR$_{nbr}$}: The effect of FaiRIR$_{nbr}$ is also shown in Tables~\ref{Tab: ML-10M} and~\ref{Tab: Amazon}. 
In all cases, the exposure bias has decreased substantially (almost reduced to zero), with a significant increase in percentage of items being adequately exposed.

\begin{table}[tb]
	\noindent
	\scriptsize
	\centering
	\begin{tabular}{|p{1cm}|p{1.2cm}||p{1.0cm}|p{1.0 cm}|p{1.0cm}||p{1.0 cm}| }
		\hline
		Algorithm &  & Over & Adequate & Under & $ExpBia$s \\
		\hline
		\multicolumn{6}{|c|}{\bf $\beta = 0.0$}\\
		\hline \hline
		& Vanilla 		& 18.04 \% & 06.62 \% & 75.34 \% & 1.28\\ \cline{2-6}
		& FaiRIR$_{concat}$	& 20.14 \% & 12.78 \% & 67.1 \% & 0.75\\ \cline{2-6}
		& FaiRIR$_{rl}$ 	& 33.01 \% & 23.71 \% & 43.28 \% & 0.18\\ \cline{2-6}
		rating-& FaiRIR$_{sim}$	& 14.60 \% & 8.70 \% & 76.70 \% & 1.2\\ \cline{2-6}
		-SVD& FaiRIR$_{nbr}$	& 0.0 \% & 100.0 \% & 0.0 \% & 0.002\\ \cline{2-6}
		\hline \hline
		& Vanilla 		& 15.73 \%& 05.67 \% & 78.60 \% & 1.22  \\ \cline{2-6}
		& FaiRIR$_{concat}$	& 31.26 \%& 19.41 \% & 49.33 \% & 0.32  \\ \cline{2-6}
		& FaiRIR$_{rl}$ 	& 35.17 \%& 26.71 \% & 38.12 \% & 0.14  \\ \cline{2-6}
		item2vec & FaiRIR$_{sim}$	& 20.65 \%& 16.77 \% & 62.57 \% & 0.81  \\ \cline{2-6}
		& FaiRIR$_{nbr}$	& 0.0 \% & 100.0 \% & 0.0 \% & 0.002  \\ 
		\hline \hline
		\multicolumn{6}{|c|}{\bf $\beta = 0.25$}\\
		\hline 
		& Vanilla 		& 15.66 \%& 05.66 \% & 78.68 \% & 1.21  \\ \cline{2-6}
		item2vec& FaiRIR$_{rl}$ 	& 32.43 \%& 22.66 \% & 44.91 \% & 0.18  \\ \cline{2-6}
		& FaiRIR$_{sim}$	& 18.90 \%& 7.67 \% & 73.43 \% & 0.91  \\ \cline{2-6}
		& FaiRIR$_{nbr}$	& 0.0 \% & 100.0 \% & 0.0 \% & 0.001  \\ 
		\hline
		\multicolumn{6}{|c|}{\bf $\beta = 0.75$}\\
		\hline 
		& Vanilla 		& 15.46 \%& 05.68 \% & 78.86 \% & 1.21  \\ \cline{2-6}
		item2vec& FaiRIR$_{rl}$ 	& 32.89 \%& 24.28 \% & 42.83 \% & 0.16  \\ \cline{2-6}
		& FaiRIR$_{sim}$	& 15.96 \%& 6.67 \% & 77.57 \% & 1.14  \\ \cline{2-6}
		& FaiRIR$_{nbr}$	& 0.0 \% & 100.0 \% & 0.0 \% & 0.0002  \\ 
		\hline
		\multicolumn{6}{|c|}{\bf $\beta = 1.00$}\\
		\hline 
		& Vanilla 		& 15.41 \%& 05.78 \% & 78.81 \% & 1.21  \\ \cline{2-6}
		item2vec& FaiRIR$_{rl}$ 	& 32.12 \%& 30.37 \% & 37.51 \% & 0.12  \\ \cline{2-6}
		& FaiRIR$_{sim}$	& 15.41 \%& 05.78 \% & 78.81 \% & 1.21  \\ \cline{2-6}
		& FaiRIR$_{nbr}$	& 0.0 \% & 100.0 \% & 0.0 \% & 0.0  \\ 
		\hline
	\end{tabular}
	\caption{{\bf \% of items that are over, adequately and under-exposed in Amazon review dataset (vanilla and intervened item2vec. For rating-SVD representative results shown for $\beta=0.0$; other results are similar to item2vec and omitted).}}
	\label{Tab: Amazon}
	\vspace{-6 mm}
\end{table}

\begin{figure}[tb]
	\centering
	\begin{subfigure}{0.32\columnwidth}
		\centering
		\includegraphics[width=\textwidth, height=2.75cm]{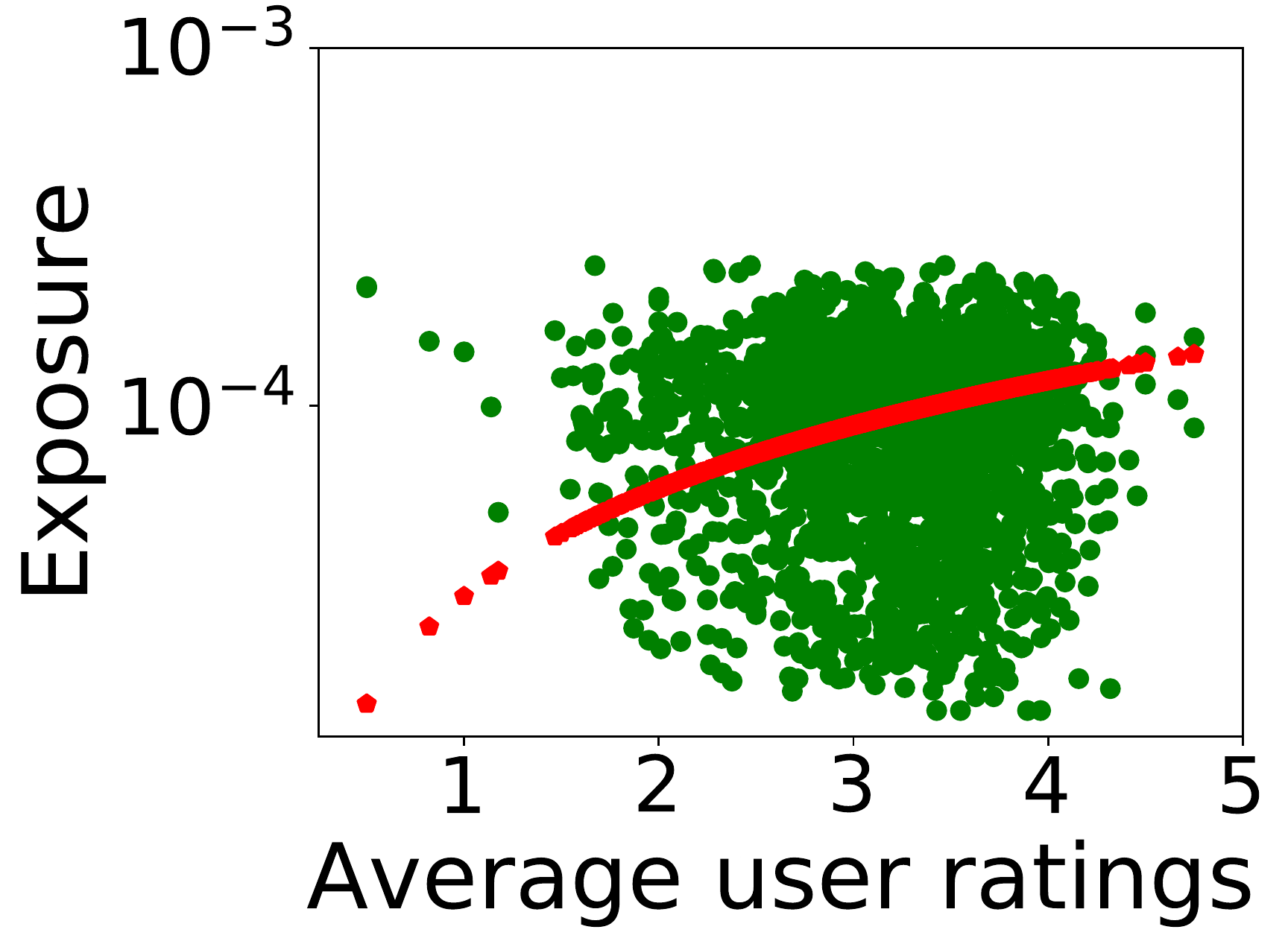}
		\caption{FaiRIR$_{rl}$}
		\label{fig: Phase1Exp}
	\end{subfigure}%
	\begin{subfigure}{0.32\columnwidth}
		\centering
		\includegraphics[width=\textwidth, height=2.75cm]{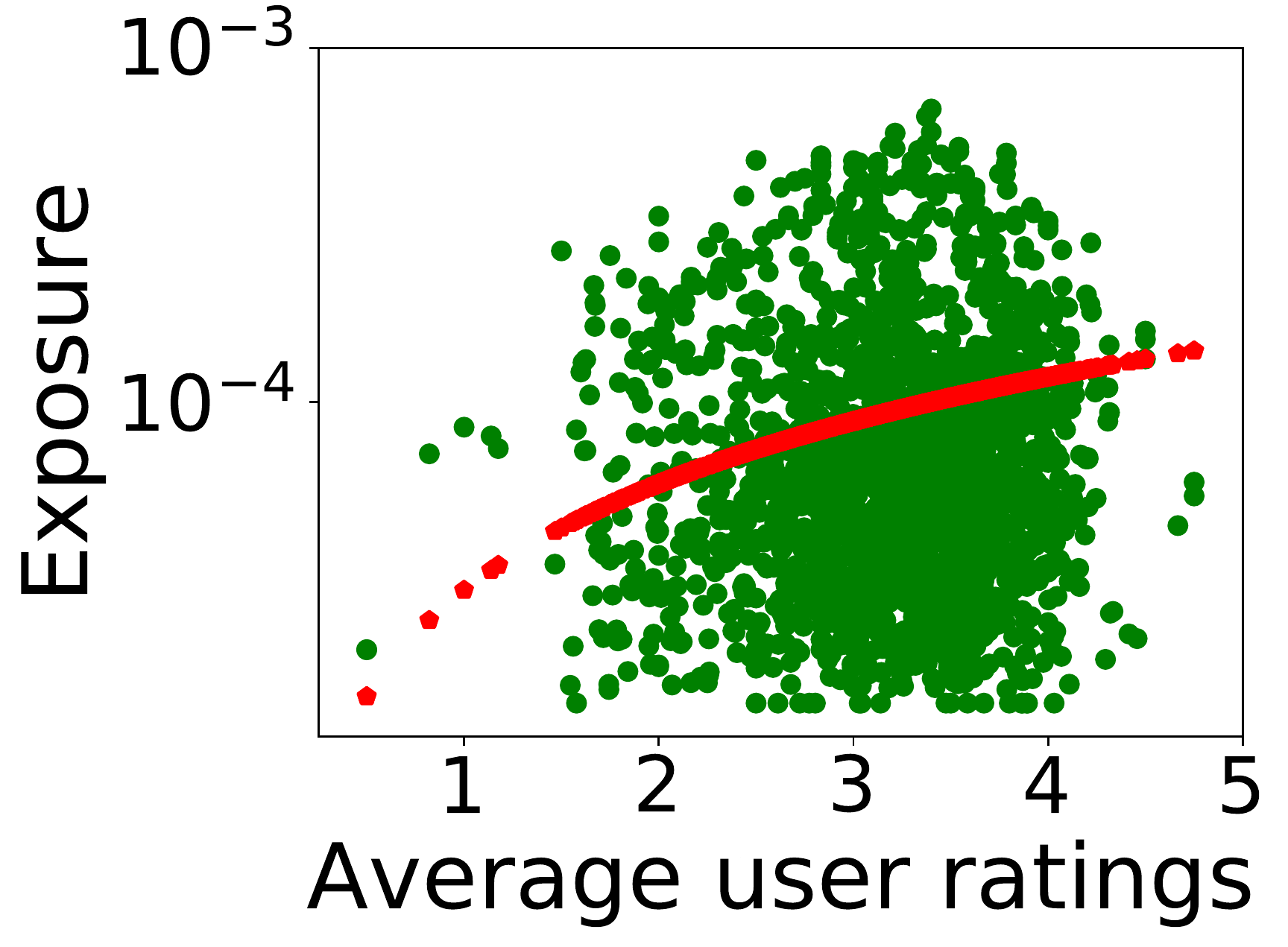}
		\caption{FaiRIR$_{sim}$}
		\label{fig: Phase2Exp}
	\end{subfigure}
	\begin{subfigure}{0.32\columnwidth}
		\centering
		\includegraphics[width=\textwidth, height=2.75cm]{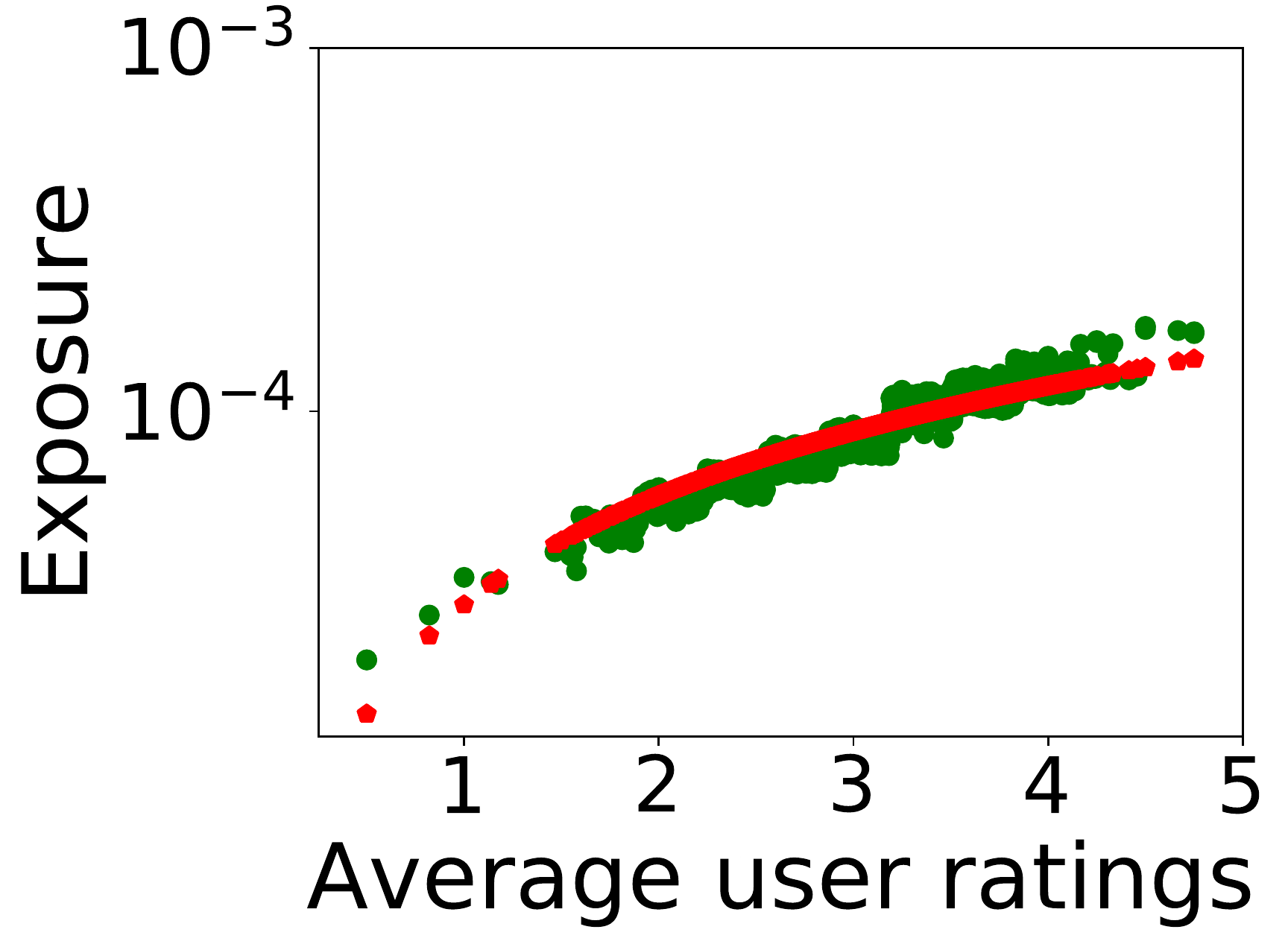}
		\caption{FaiRIR$_{nbr}$}
		\label{fig: Phase3Exp}
	\end{subfigure}
	\vspace*{-2 mm}
	\caption{{\bf (color online) Scatter plot of log of observed exposure and average user rating of movies in MovieLens dataset after different interventions on the rating-SVD. The red scatter shows their desired exposure with $\beta = 0.0$.} 
	}
	\label{fig: FairExp}
	\vspace*{-3 mm}
\end{figure}

\noindent Figure~\ref{fig: FairExp} shows scatter plots for the three interventions on the MovieLens dataset; each plot shows log of observed exposure on $y$-axis and desired exposure ($\beta = 0.0$) on $x$-axis. From these figures as well, it is evident that the distribution of observed exposure is closest to that of desired exposure for FaiRIR$_{nbr}$ (Figure~\ref{fig: FairExp}(c)).

\noindent
\textbf{Analysis on multiple desired exposure distribution}: We also analyze our proposed interventions with different desired exposure distributions, by varying $\beta$ in the range $[0,1]$. The results for Amazon dataset are shown in Table~\ref{Tab: Amazon} (similar results are obtained for the MovieLens dataset, not shown for brevity). We see that irrespective of the $\beta$ values, FaiRIR$_{rl}$ and FaiRIR$_{nbr}$ are very effective in mitigating the exposure bias. However, with increase in $\beta$, the effectiveness of FaiRIR$_{sim}$ reduces. 
The reason being, as $\beta$ approaches $1.0$ (i.e., uniform desired exposure distribution), the \textit{`desired exposure based similarity'} part in equation~\ref{eqn: p2} reduces to $1.0$ and $sim(x_i, x_j)$ becomes only the relatedness similarity (cosine similarity).

Overall, FaiRIR$_{nbr}$ is seen to be most effective in reducing exposure bias, probably due to the following reason. 
While FaiRIR$_{rl}$ and FaiRIR$_{sim}$ attempt to reduce exposure bias indirectly by altering the representation learning / similarity computation in the latent space, FaiRIR$_{nbr}$ directly controls the number of other items from which a particular item $i$ is recommended (the desired number of recommendations for $i$), which specifically ensures that $i$ gets an exposure close to its desired exposure.

\noindent
\textbf{Summary:} We have shown that the proposed methods 
are successful in mitigating the induced exposure bias significantly. While the performance of  FaiRIR$_{rl}$ and FaiRIR$_{nbr}$ show significant improvement across both the datasets and RIR algorithms, the improvement is less stable for FaiRIR$_{sim}$.

The next natural question to investigate is whether this control of exposure bias comes at a cost of loss in relatedness of the recommendations.

\subsection{Preserving recommendation relatedness} \label{sub:expt-reco-relatedness}

Mitigating bias due to algorithms usually
incur an associated cost in terms of drop in performance (e.g., drop in accuracy in fair classifiers 
\cite{zafar2017fairness})
. 
In the context of RIRs, the relatedness of the recommendation is the prime objective of the algorithm. 
Hence, in this experiment we check the incurred loss in relatedness due to our proposed FaiRIR interventions.

\noindent \textbf{Genre / Category based similarity}: Intuitively, relatedness of recommendations will be high if the recommended items are `similar'/`related' to the source item. The measure of relatedness is often domain-dependent. For instance, in the domain of movies, every movie has a set of one or more genres.
We measure the similarity of a source movie and a recommended movie by their {\it genre overlap} with respect to the source movie, i.e., by the fraction of genres preserved by the recommended movie as compared to that of the source movie.
For instance, let the movie {\it Gladiator} be recommended from the movie {\it Avatar} by an algorithm. As per the IMDb website, {\it Avatar} has genres \{Action, Adventure, Fantasy\} and {\it Gladiator} has genres \{Action, Adventure, Drama\}. Thus, the similarity between the movies is $\frac{2}{3}$.
Similarly, every item in the Amazon dataset has an associated set of {\it categories}, which we used in an identical fashion to compute relatedness of two Amazon items.
Now, we compute the relatedness of RIRs generated by an algorithm as follows. 
For each pair of items $(i,j)$ where $j$ has been recommended for $i$ by the algorithm, we compute the genre (category) overlap between $i$ and $j$, and then take the mean across all such pairs.

\begin{table}[tb]
	\noindent
	\scriptsize
	\centering
	\begin{tabular}{|p{1cm}|p{1cm}|c|c|c|}
		\hline
		Algorithm &  & Genre overlap & Like overlap & Relevance score \\
		\hline
		& Vanilla 		& 0.53 & 0.64 & 3.73\\ \cline{2-5}
		& FaiRIR$_{rl}$	& 0.34 & 0.56 & 2.69\\ \cline{2-5}
		rating-	  & FaiRIR$_{sim}$	& 0.44 & 0.64 & 3.63\\ \cline{2-5}
		-SVD& FaiRIR$_{nbr}$	& 0.44 & 0.63 & 3.62\\
		\hline \hline
		& Vanilla 		& 0.37 & 0.56 & 2.79\\ \cline{2-5}
		& FaiRIR$_{rl}$	& 0.27 & 0.52 & 2.54\\ \cline{2-5}
		item2vec & FaiRIR$_{sim}$	& 0.35 & 0.59 & 2.77 \\ \cline{2-5}
		& FaiRIR$_{nbr}$	& 0.34 & 0.59 & 2.76\\
		\hline
		
	\end{tabular}	
	\caption{{\bf Relatedness of recommendations generated 
			over the MovieLens dataset: (i)~mean genre overlap of recommended \& source item, (ii)~Like overlap between users who liked the source \& recommended items, (iii)~relevance score according to AMT survey.}}
	\label{Tab:relatedness-of-recs}
	\vspace{-4 mm}
\end{table}

\begin{table}[tb]
	\noindent
	\scriptsize
	\centering
	\begin{tabular}{|p{0.85cm}|p{1 cm}|c|c|c|c| }
		\hline
		Algorithm & & $\beta = 0.0$ & $\beta = 0.25$ & $\beta = 0.75$ & $\beta = 1.0$ \\
		\hline
		& Vanilla 		& 0.46 & 0.46 & 0.46 & 0.46 \\ \cline{2-6}
		& FaiRIR$_{rl}$	& 0.41 & 0.40 & 0.41 & 0.40\\ 
		\cline{2-6}
		rating-& FaiRIR$_{sim}$	& 0.45 & 0.45 & 0.46 & 0.46\\ \cline{2-6}
		-SVD & FaiRIR$_{nbr}$	& 0.44 & 0.44 & 0.45 & 0.45 \\
		\hline \hline
		& Vanilla 		& 0.50 & 0.50 & 0.50 & 0.50 \\ \cline{2-6}
		item2vec & FaiRIR$_{rl}$ 	& 0.41 & 0.41 & 0.41 & 0.40  \\ \cline{2-6}
		& FaiRIR$_{sim}$	& 0.48 & 0.48 & 0.49 & 0.50\\ \cline{2-6}
		& FaiRIR$_{nbr}$	& 0.46 & 0.46 & 0.47 & 0.47 \\
		\hline
	\end{tabular}	
	\caption{{\bf Relatedness of recommendations (mean category overlap between source and recommended items) over Amazon dataset, shown for different desired 
			distributions.}}
	\label{Tab:relatedness-of-recs-Amazon}
	\vspace{-6 mm}
\end{table}

\noindent \textbf{Results:} The `Genre overlap' column of Table~\ref{Tab:relatedness-of-recs} and Table~\ref{Tab:relatedness-of-recs-Amazon} show the genre/category overlap for the various RIR algorithms, for the MovieLens and Amazon datasets respectively. 
We observe decrements in the average genre overlap for all the interventions (as compared to the original algorithms), which is the expected cost of minimizing exposure bias. 
The reduction is severe in case of FaiRIR$_{rl}$ only, whereas it is not so severe in case of FaiRIR$_{sim}$ and FaiRIR$_{nbr}$. 
Note that, FaiRIR$_{rl}$ is an \textit{application-agnostic} methodology, where the main objective is to learn fair representations, while the other two interventions are \textit{application-specific}. 
Hence, the foregoing observation is in line with the findings in fairness literature, wherein application-agnostic approaches tend to incur higher losses in performance~\cite{lahoti2019ifair}.

\noindent
\textbf{Analysis on multiple desired exposure distribution}: Table~\ref{Tab:relatedness-of-recs-Amazon} shows the category overlap for various desired exposure distributions (various values of $\beta$) for the Amazon product review dataset. The interpretation of the results is the same as discussed above. The performance of all the FaiRIR algorithms in preserving relatedness is pretty much stable. 
However, since FaiRIR$_{sim}$ reduces to vanilla RIR algorithms as $\beta$ approaches $1.0$, it does not incur any additional loss in 
relatedness.


\vspace{-1 mm}
\subsection{Judging overall utility of recommendations} 
\label{sec: user_survey}
Considering that the ultimate objective of RIR algorithms is to satisfy human users, we conduct the following two evaluations for the proposed algorithms -- 
(1)~We measure the mean overlap of common users who liked both a source item and a recommended item, and 
(2)~We conduct a user survey for judging the utility (relevance) of the recommendations generated by various algorithms on the MovieLens dataset.

\vspace{1mm}
\noindent 
\textbf{(1) Users' propensity to like the recommended items}:
When a source item $i$ recommends an item $j$,
intuitively, the utility of the recommendation is high if the recommended item $j$ is `liked' by most users who had also liked the source item $i$. 
To this end, we consider a user to have liked a given movie / item if (s)he gives a rating/score of more than 3.5 (out of 5) i.e., a score of more than 70\%. Our choice is influenced by studies claiming 3.5--4.5 (out of 5) being the sweet spot~\cite{WomplyHow, MetacriticHow} in five star rating systems. 
Note that, since all users who had liked $i$ need not consume or rate item $j$, we need to consider only those users who liked $i$ and rated the recommended item $j$. 
For the item $i$, let $R_i$ and $L_i$ be the set of users who rated and liked the item respectively ($L_i$ is a subset of $R_i$). 
For a source item $i$ and a recommended item $j$ (from $i$), we evaluate the \textbf{like overlap} as $\frac{|L_i \cap L_j|}{|L_i \cap R_j|}$, i.e., out of all users who liked item $i$ and rated item $j$, what fraction of users actually liked (rated highly) $j$.
Specifically, for each pair of items $(i,j)$ where $j$ has been recommended for $i$, we compute the like overlap between $i$ and $j$, and then take the mean across all such pairs.

\vspace{1 mm}
\noindent \textbf{Results:}  Table~\ref{Tab:relatedness-of-recs} (`Like overlap' column) and Table~\ref{Tab:UserSatisfaction-of-recs-Amazon} (all columns, for different $\beta$ values) show the mean {\it like overlap} for the various RIR algorithms, for the MovieLens and Amazon datasets respectively. 
For the MovieLens data (Table~\ref{Tab:relatedness-of-recs} last column), while using rating-SVD, the average {\it like overlap} reduces for FaiRIR$_{rl}$ (as compared to the original algorithm); however, there is practically no reduction in like overlap for FaiRIR$_{sim}$ and FaiRIR$_{nbr}$.  
Interestingly, while using item2vec, {\it FaiRIR$_{sim}$ and FaiRIR$_{nbr}$} ({\it like overlap = 0.59}) {\it have outperformed even the original algorithm} ({\it like overlap = 0.56}). 
For Amazon dataset (Table~\ref{Tab:UserSatisfaction-of-recs-Amazon}), for both rating-SVD and item2vec algorithms, all the FaiRIR algorithms have performed comparably or even better than the original algorithms. We repeated this experiment with 4 and 4.5 thresholds for {\it like overlap} and found qualitatively similar observations (omitted for brevity). These results show the efficacy of the proposed approaches to preserve utility of the recommendation to end users. However, one can argue that such high scores might be an artifact of the underlying desiredness assumption while designing the algorithm esp. when $\beta = 0$ i.e., a completely meritocratic distribution. Hence, analysis on different $\beta$ values are of utmost importance.

\noindent
\textbf{Analysis on multiple desired exposure distribution}: Table~\ref{Tab:UserSatisfaction-of-recs-Amazon} shows the {\it like overlap} for various desired exposure distributions (i.e., for various values of $\beta$) for the Amazon dataset. 
We see that the performance of all the FaiRIR interventions in preserving utility toward users is very stable and is agnostic of the underlying desired distribution (similar results for MovieLens dataset are omitted for brevity).

\vspace{2mm}
\noindent \textbf{(2) A user survey to judge recommendations}: 
We recruited human workers to assess the relevance of recommendations, via the Amazon Mechanical Turk (AMT) platform. 
We used `AMT master workers' who are known to perform such tasks meticulously. 
To ensure reliable judgments, it is important that the annotators are likely to be familiar with the items whose recommendations they are being tasked to judge.
Hence, we chose to perform this survey with the MovieLens dataset, since workers are much more likely to be familiar with popular movies, than with cellphones and accessories available on Amazon.
We considered movies in the top 5\% most popular movies (based on the number of ratings) in MovieLens dataset as our source movies. 
To further guarantee the reliability of the judgments, we also provided the annotators with the link to the IMDb information page of each movie, which contains all metadata about the movie along with a snippet and its trailer. 
The annotators were asked to browse through the IMDb page for a movie if they were not familiar with it.

For a particular source movie $x$, we generated top-$5$ recommendations using various algorithms on the MovieLens dataset.
For each movie $y$ recommended for $x$ (in top-$5$) by some algorithm, we asked a worker -- ``If your friend likes the movie $x$, how likely are you to recommend movie $y$?''. 
A worker could answer this question on a Likert scale of $[1,5]$ with response $1$ representing ``very unlikely'' and response $5$ representing ``very likely''.
The recommendations were anonymized, i.e., the workers were {\it not} mentioned 
about the source algorithm of each recommendations.

We collected responses for $100$ different source movies and approximately $1,550$ distinct pairs of source and recommended pairs of movies. Each pair was evaluated by at least $10$ AMT workers, and we considered the average score over all these AMT workers as the utility/relevance score for this pair. 
The mean relevance for a RIR algorithm is computed as the average relevance over all the recommended pairs generated by it that were evaluated.

\noindent
\textbf{Results:} 
Each of the RIR algorithms scored based on the mean of the relevance scores of different items it recommended 
for a given item. 
Table~\ref{Tab:relatedness-of-recs} (Relevance Score column) shows the mean relevance scores for different RIR algorithms.

For both rating-SVD and item2vec, FaiRIR$_{sim}$ and FaiRIR$_{nbr}$ preserve a high degree of relevance in their recommendations. For instance, the mean relevance score of the original rating-SVD algorithm is $3.73$ (out of $5.0$), while that of FaiRIR$_{sim}$ and FaiRIR$_{nbr}$ are $3.63$ and $3.62$ respectively. The drop in mean relevance score is even lesser in case of item2vec.
To further substantiate the results, we 
performed {\it Student's T-test} on the samples of mean relevance scores, as obtained from the user survey. 
We found the drop in mean relevance score to be statistically significant only for FaiRIR$_{rl}$ (as compared to the original algorithm); for FaiRIR$_{sim}$ and FaiRIR$_{nbr}$, the drop 
was {\it not} statistically significant. 
In fact, for almost 40\% of the recommendations, the 
FaiRIR variants have {\it higher} relevance scores than the vanilla algorithms. 

\begin{table}[tb]
	\noindent
	\scriptsize
	\centering
	\begin{tabular}{|p{1cm}|p{1 cm}|c|c|c|c| }
		\hline
		Algorithm & & $\beta = 0.0$ & $\beta = 0.25$ & $\beta = 0.75$ & $\beta = 1.0$ \\
		\hline
		& Vanilla 		& 0.91 & 0.91 & 0.91 & 0.91 \\ \cline{2-6}
		& FaiRIR$_{rl}$ 	& 0.90 & 0.89 & 0.90 & 0.89\\ 
		\cline{2-6}
		rating-& FaiRIR$_{sim}$ 	& 0.91 & 0.91 & 0.91 & 0.91\\ \cline{2-6}
		-SVD & FaiRIR$_{nbr}$	& 0.92 & 0.91 & 0.91 & 0.91 \\
		\hline \hline
		& Vanilla 		& 0.81 & 0.81 & 0.81 & 0.81 \\ \cline{2-6}
		item2vec & FaiRIR$_{rl}$ 	& 0.87 & 0.84 & 0.83 & 0.86  \\ \cline{2-6}
		& FaiRIR$_{sim}$ & 0.85 & 0.84 & 0.82 & 0.81\\ \cline{2-6}
		& FaiRIR$_{nbr}$& 0.82 & 0.81 & 0.81 & 0.81 \\
		\hline
		
	\end{tabular}	
	\caption{{\bf Mean like overlap (overlap of users who liked the recommended \& source item) of the recommendations generated by different algorithms over Amazon dataset, shown for different desired exposure distributions.}}
	\label{Tab:UserSatisfaction-of-recs-Amazon}
	\vspace{-6 mm}
\end{table}

Note that, we also attempted to combine all the interventions, i.e., first learn fair representations, then fairly 
compute similarity, and finally select the neighbors as shown in Algorithm~\ref{Algo}. The results for the combined approach were dominated by the effects of FaiRIR$_{nbr}$.
Hence, we have not reported results of the combined approach for brevity.

\vspace{1mm}
\noindent
{\bf Summary:}  
We investigated whether the mitigation of exposure bias by FairIR comes at the cost of degradation in the relatedness of the recommendations.
Specifically we evaluated -- 
(i)~the preservation of relatedness of the recommendation through \textit{genre / category overlap}, and 
(ii)~the utility of the recommendation through {\it liking overlap} metric and a user-survey on AMT. 
Across all evaluations, we consistently observe that our proposed mechanisms (especially FaiRIR$_{sim}$ and FaiRIR$_{nbr}$) successfully mitigate exposure bias without sacrificing much on the relatedness of recommendations.

	\section{Conclusion}

In this paper, we considered the impact of related item recommendations (RIRs)
on the exposure of items. We show that existing RIRs induce exposure bias by not considering any notion of the desired exposure of items. 
Although there can be alternate ways to estimate the exposure of different items, we believe that the qualitative results would remain unchanged. 
We further proposed a novel suit of algorithms (\textbf{FaiRIR}), which can reduce the exposure bias, while maintaining the effectiveness of recommendations. 
Note that, though the experiments in this paper are conducted on two datasets (from movie and e-commerce domains)
for proof of concept, the proposed algorithms are applicable to any other domain, including job recommendation sites, and others.

In this work, we considered RIRs that only utilize the relatedness with respect to one source item. 
Many platforms provide personalized recommendations to users, where multiple items viewed by a target user are considered. We plan to study the effects of personalization on exposure bias in future.


	\section*{Acknowledgment}
	The authors thank the anonymous reviewers and the Associate Editor whose comments helped to improve the paper. This research is supported in part by (1)~a grant from the Max Planck Society through a Max Planck Partner Group at IIT Kharagpur, and (2)~a European Research Council (ERC) Advanced Grant for the project ``Foundations for Fair Social Computing", funded under the European Union's Horizon 2020 Framework Programme (grant agreement no. 789373). Additionally, A. Dash is supported by a fellowship from Tata Consultancy Services.
	
	\balance
	\bibliographystyle{IEEEtran}
	\bibliography{Main}

\end{document}